\documentclass[iop, apj, natbib, numberedappendix]{emulateapj}
\usepackage{apjfonts} 
\usepackage{amsmath}
\usepackage{lmodern}
\usepackage{amssymb}
\usepackage{color}
\usepackage{multirow}
\usepackage{rotating}
\usepackage{booktabs}
\usepackage{times}
\usepackage{natbib}

\newcommand{\Msun}{\ensuremath{M_{\odot}}}

\newcommand{\zp}{\ensuremath{z^\prime}}

\def\farcs{\hbox{$.\!\!^{\prime\prime}$}}

\newcommand{\nbest}{\textit{NB816}}

\newcommand{\nbnto}{\textit{NB921}}

\newcommand{\ha}{H$\alpha$}
\newcommand{\hb}{H$\beta$}
\newcommand{\hg}{H$\gamma$}
\newcommand{\hi}{H\,{\sc i}}	
	
\newcommand{\oii}{[O\,{\sc ii}]}	
\newcommand{\oiii}{[O\,{\sc iii}]}
\newcommand{\nev}{[Ne\,{\sc v}]}
\newcommand{\mgii}{Mg\,{\sc ii}}
\newcommand{\feii}{Fe\,{\sc ii}}
\newcommand{\heii}{He\,{\sc ii}}
\newcommand{\feiis}{Fe\,{\sc ii}$^{\ast}$}
\newcommand{\nai}{Na\,{\sc i}}

\newcommand{\hii}{H\,{\sc ii}}
\newcommand{\nii}{[N\,{\sc ii}]}

\newcommand{\oiib}{O\,{\sc ii}B}	
\newcommand{\oiiib}{O\,{\sc iii}B}	
\newcommand{\hab}{H{\sc A}B}	

\newcommand{\rz}{\textit{Rz}}

\newcommand{\obone}{O\,{\sc iii}B-1}
\newcommand{\obtwo}{O\,{\sc iii}B-2}
\newcommand{\obthree}{O\,{\sc iii}B-3}
\newcommand{\obfour}{O\,{\sc iii}B-4}

\newcommand{\kms}{km\,s$^{-1}$}
\newcommand{\cms}{cm\,s$^{-1}$}
\newcommand{\ergs}{erg\,s$^{-1}$}

\newcommand{\ergscm}{erg\,s$^{-1}$\,cm$^{-2}$}
\newcommand{\ergscmarcsec}{erg\,s$^{-1}$\,cm$^{-2}$\,arcsec$^{-2}$}

\slugcomment{Accepted for Publication in The Astrophysical Journal}

\shorttitle{Spatially Extended \oiii\ Sources}

\shortauthors{Yuma et al.}

\begin{document}

\title{A Giant Green Pea Identified in the Spectroscopy of Spatially Extended \oiii\ Sources
}

\author{Suraphong Yuma$^{1}$}
\author{Masami Ouchi$^{2}$}
\author{Seiji Fujimoto$^{2,3}$}
\author{Takashi Kojima$^{2,4}$}
\author{Yuma Sugahara$^{2,4}$}

\affil{$^1$Department of Physics, Faculty of Science, Mahidol University, Bangkok 10400, Thailand; suraphong.yum@mahidol.ac.th}
\affil{$^2$Institute for Cosmic Ray Research, The University of Tokyo, Kashiwa-no-ha, Kashiwa 277-8582, Japan}
\affil{$^3$Department of Astronomy, University of Tokyo, Hongo, Bunkyo-ku, Tokyo 113-0033, Japan}
\affil{$^4$Department of Physics, University of Tokyo, Hongo, Bunkyo-ku, Tokyo 113-0033, Japan}


\begin{abstract}
We present the results of the deep Subaru/FOCAS and Keck/MOSFIRE spectroscopy 
for four spatially extended \oiii$\lambda\lambda4959,5007$ sources, 
dubbed \oiii\ blobs, at $z=0.6-0.8$ that are originally pinpointed 
by large-area Subaru imaging surveys. The line diagnostics of the rest-frame 
optical lines suggests that only one \oiii\ blob, OIIIB-3, 
presents an AGN signature, indicating that hot gas of the rest of the \oiii\ blobs is 
heated by star formation. One of such star-forming \oiii\ blobs, OIIIB-4, at $z=0.838$ 
has an \oiii\ equivalent width of $845\pm27$ \AA\ and 
an \oiii\ to \oii$\lambda\lambda3726,3729$ ratio of \oiii/\oii$ = 6.5 \pm 2.7$ 
that are as high as those of typical green peas (Cardamone et al. 2009). 
The spatially resolved spectrum of OIIIB-4 
shows \oiii/\oii$=5-10$ over $14$ kpc in the entire large \oiii\ extended regions of OIIIB-4,
unlike the known green peas whose strong \oiii\ emission region is compact. 
Moreover, OIIIB-4 presents no high-ionization emission lines unlike {\it green beans} that have extended \oiii\ emission with a type-2 AGN. 
OIIIB-4 is thus a {\it giant} green pea, which is a low stellar mass ($7\times10^7$\Msun) 
galaxy with a very high specific star-formation rate (sSFR $=2\times10^2$ Gyr$^{-1}$), 
a high ionization parameter ($q_{ion}\sim 3\times10^8$\cms), 
and a low metallicity 
similar to those of green peas. Neither an AGN-light echo nor a fast radiative shock likely takes place due to
the line diagnostics for spatially-resolved components of OIIIB-4 and no detections of 
\heii$\lambda4686$ or \nev$\lambda3346,3426$ lines that are fast-radiative shock signatures. 
There is a possibility that
the spatially-extended \oiii\ emission of OIIIB-4 is originated from outflowing gas produced 
by the intense star formation in a density-bounded ionization state.

\end{abstract}


\section{Introduction}\label{sec:intro}

A galaxy is not a closed box that can evolve itself without exchanging materials with 
the environment. 
The accretion of cooled gas onto the galaxy is important for the galaxy to form new stars in 
the $\Lambda$ Cold Dark Matter ($\Lambda$CDM) model. 
However, the inflow alone would cause an overestimation of star formation at both low-mass and high-mass ends  \citep[e.g.,][]{bell03, mutch13}. 
The feedback mechanism involving gas outflows becomes a default tool to resolve the discrepancy in 
hydrodynamic simulations \citep[e.g.,][]{benson03, somerville08, oppenheimer10, vandevoort11}. 
The evolutional state of a galaxy is thus thought to depend on the balance of the gas flows in and out of the galaxy 
\citep[e.g.,][]{lilly13}. 
In addition, the galactic-scale outflow is considered a solution for various observational phenomena including 
regulating the main sequence of galaxies on the mass-SFR (star formation rate) diagram \citep[e.g.,][]{noeske07} 
and the mass-metallicity relation  \citep[e.g.,][]{tremonti04}, and enriching the chemical abundance of 
interstellar medium (ISM) and intergalactic medium \citep[IGM; e.g.,][]{martin05, rupke05a, rupke05b, weiner09, coil11}. 

Outflows have been massively studied in several types of galaxies ranging from  
normal star-forming galaxies \citep[e.g.,][]{weiner09, steidel10, erb12, martin12, bradshaw13, rubin14} 
and submillimeter galaxies \citep[SMG; e.g., ][]{alexander10} to more extreme systems like 
radio galaxies \citep[e.g.,][]{nesvadba08, liu13}, 
ultra luminous infrared galaxies \citep[ULIRGs, e.g.,][]{heckman90, martin05, rupke05a, rupke05b, soto12}, and 
active galactic nuclei \citep[AGNs; e.g.,][]{cicone14, cheung16, rupke17}. 
Outflows are found to be ubiquitous in galaxies with the SFR surface density larger than 
$\sim0.1$ \Msun\,yr$^{-1}$\,kpc$^{-2}$ and become stronger in more massive star-forming galaxies 
with higher SFRs \citep{heckman00, martin05, weiner09, kornei12, martin12}. 
In the case of AGNs, the outflows strongly correlate with the black hole mass ($M_{\rm BH}$) in that 
the mass outflow rate increases with increasing $M_{\rm BH}$ \citep{rupke17}. 
An active supermassive black hole alone could fuel the large-scale outflow even in a low-luminosity AGN \citep{cheung16}. 

Most of these studies are based mainly on the optical/near-infrared 
spectroscopic observations of the blueshifted interstellar absorption lines 
such as \nai$\lambda\lambda5890, 5896$, \mgii$\lambda\lambda2796, 2803$, \feii$\lambda2374$, 
and \feii$\lambda2383$ to indicate the outflow signature. 
No systematic search had been done by using only the imaging data until \cite{yuma13} proposed 
to use the narrowband technique to systematically identify the gas outflowing galaxies 
by selecting star-forming galaxies with the strong \oii$\lambda\lambda3726,3729$ emission line 
spatially extended over 30 kpc beyond the stellar components. 
They called this \oii\ extended object ``\oii\ blob" or in short ``\oiib." 
The spatial extension of a metal line like the \oii\ emission line that is redshifted and falls into the narrowband image 
might indicate a large-scale outflow beyond the galaxy rather than an evidence of a gas inflow 
from the metal-poor IGM \citep[e.g.,][]{aguirre08, fumagalli11}. 
\cite{yuma13} successfully identified twelve \oiib s at $z\sim1.2$, one of which is classified as an obscured AGN. 
The others are potentially normal star-forming galaxies as they are not detected in X-ray or radio 1.4 GHz 
wavelengths. 
The hypothesis that the extended \oii\ emission line may represent the outflow from the \oii\ blobs  
is spectroscopically confirmed by the traditional method of detecting blueshifted interstellar absorption lines 
\citep{yuma13, harikane14}. 
The systematic search was then expanded toward lower and higher redshifts; i.e., $z=0.1-1.5$ \citep[][]{yuma17}. 
In \cite{yuma17}, the spatially extended emission lines include \oii$\lambda\lambda3726,3729$, 
\oiii$\lambda5007$, and \ha$\lambda6563$. 
The resulting sample is named accordingly as an \oii\ blob (\oiib), an \oiii\ blob (\oiiib), and an \ha\ blob (\hab). 
They found 77 blobs at $z=0.4-1.5$ in total, eight of which are likely to be powered by AGNs. 
It is suggested that the blobs are mostly normal star-forming galaxies. 
It is of interest to investigate an energy source for the extended emission lines of the blobs 
that are identified as normal star-forming galaxies. 

Galaxies with large extension of emission lines have recently been discovered by many studies. 
\cite{lin17} discovered a giant \ha\ blob with the \ha\ extent of $3-4$ kpc in radius. 
The spatial extension of the \oiii\ emission line is also found \citep{brammer13, sun17}. 
An AGN is thought to be a primary source of energy to produce the large-scale outflow in 
both the \ha\ blob discovered by \cite{lin17} and the \oiii\ blobs studied by \cite{sun17}. 
The latter work suggested that the size of \oiii\ extension strongly correlates with the AGN luminosity. 
\cite{sun17} also argued that there is no AGN luminosity threshold for launching the outflow, 
which is consistent with \cite{cheung16}. 
It means that it is common to see the extended emission line in all AGNs 
regardless of their luminosities. 

Although there are certain detailed studies about the extended emission lines 
in the normal star-forming galaxies \citep[e.g.,][]{genzel11, newman12b, newman12a}, 
the physical mechanism responsible for the spatially extended emission lines 
beyond the stellar component of the galaxies is still not well understood. 
To date, many studies about the ISM of galaxies are conducted with the integral field unit (IFU) observations \citep[e.g.,][]{forster09, forster14, lin17, sun17}. 
This three-dimensional (3D) imaging spectrograph enables researchers to investigate the spectra of the entire galaxies simultaneously with 2D imaging. 
In this paper, instead of using the IFU, 
we take a simple step by observing the \oiii\ blobs at $z=0.63$ and $z=0.83$ that show 
no AGN signature in X-ray and radio wavelengths 
originally selected by \cite{yuma17} with a spectrograph in the multi-object spectroscopy mode. 
By designing the slit direction to cover the longest extension of the emission lines of the blobs, we are able to determine the emission-line ratios and examine the physical properties of the extended component of the emission lines along the slit direction. 

In section \ref{sec:target}, we explain about the \oiii-blob targets for the spectroscopic observations. 
Section \ref{sec:data} describes details of both optical and near-infrared 
observations and data reduction processes.
We then report the results of systematic redshifts, outflow signature, AGN contribution, 
and properties of the ISM including the radial profile of the \oiii\ blobs in Section \ref{sec:results}. 
We also report the discovery of a giant green pea in this section. 
In section \ref{sec:discuss}, we discuss the plausible scenarios that could be responsible for  
the spatial extension of the emission lines seen in the \oiii\ blobs. 
The last section, Section \ref{sec:summary}, is the summary of all our findings. 
Throughout this paper, we adopt the standard $\Lambda$CDM cosmology with 
$H_0=70$ \kms\,Mpc$^{-1}$, $\Omega_m=0.3$, and $\Omega_\lambda=0.7$. 
All magnitudes are given in the AB system \citep{oke83}. 

\begin{figure}
	\centering
	\includegraphics[width=0.45\textwidth]{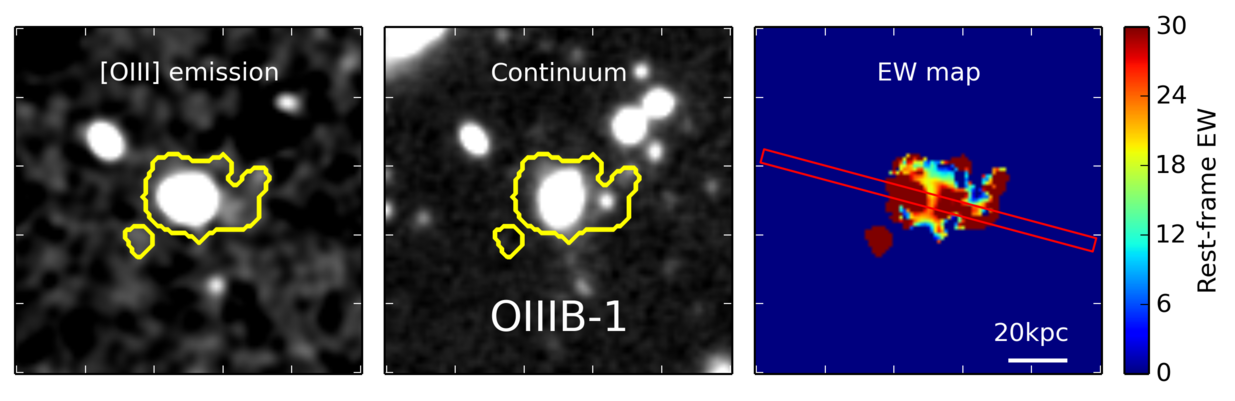}\\
	\includegraphics[width=0.45\textwidth]{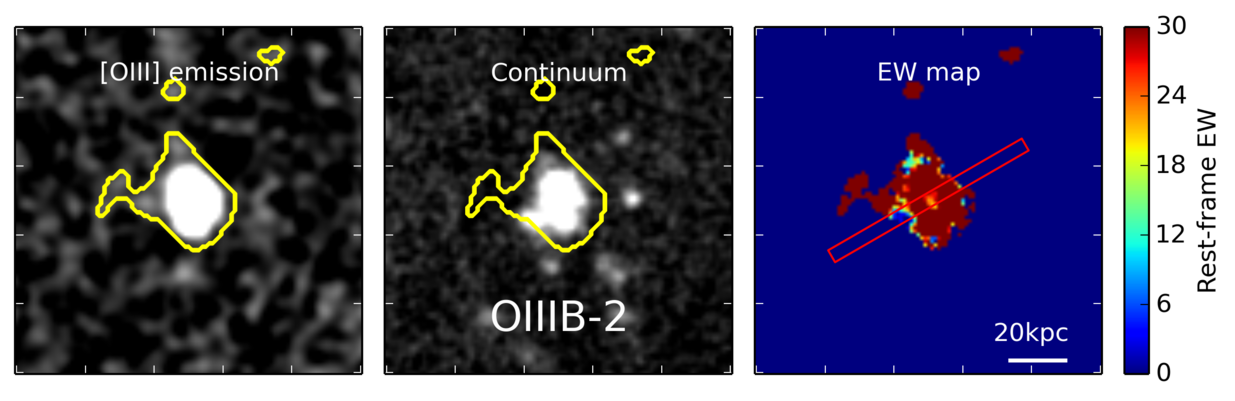}\\
	\includegraphics[width=0.45\textwidth]{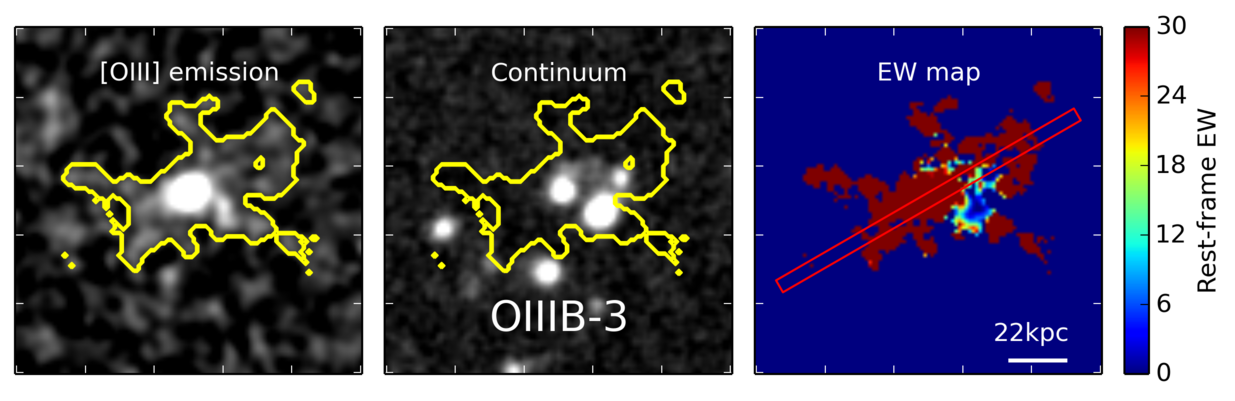}\\
	\includegraphics[width=0.45\textwidth]{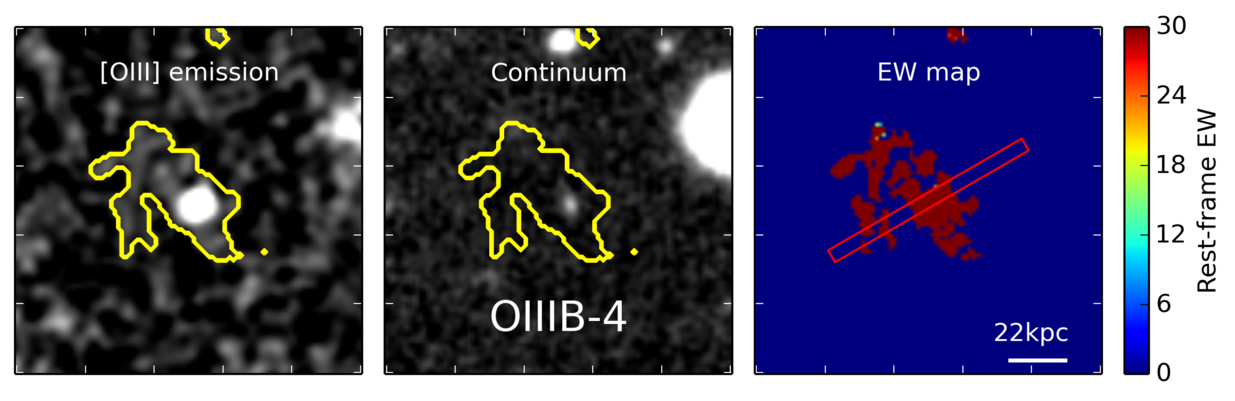}\\
	\caption{From left to right: the emission-line ($NB-BB$), stellar continuum, and 
	equivalent-width maps of 4 \oiii\ blobs at $z=0.63$ and $z=0.83$, respectively. 
	The emission-line images are \nbest-\rz\ and \nbnto-\zp\ for the blobs 
	at $z=0.63$ and $z=0.83$, respectively. 
	\rz\ is defined as {\it Rz $\equiv$ (R+2\zp)/3}. 
	IDs are indicated in the middle panel. 
	North is up and East is to the left. 
	Yellow contours in the left and middle panels indicate the isophotal area of the 
	emission-line flux measured down to $1.2\times10^{-18}$ \ergscmarcsec. 
	The color bar in the right panel corresponds to the rest-frame equivalent width per pixel.
	The magenta rectangle shows the slit direction 
	that we used to observe the spectrum of the \oiii\ blobs with Subaru/FOCAS. 
	}
	\label{fig_spec_target}
\end{figure}
\begin{deluxetable*}{l ll c cc}[!h]
\tabletypesize{\small}
\tablewidth{0pt}
\tablecolumns{11}
\tablecaption{Photometric properties of \oiii\ blobs at $z=0.63$ and $z=0.83$ for spectroscopic follow-up\label{tab_blob_sample}}
\tablewidth{0pt}
\tablehead{
\multicolumn{1}{c}{Object Name} &
\multicolumn{1}{c}{$\alpha$ (J2000)} &
\multicolumn{1}{c}{$\delta$ (J2000)} &
\multicolumn{1}{c}{$NB - BB$} & 
\multicolumn{1}{c}{$L$(\oiii)\tablenotemark{a}} & 
\multicolumn{1}{c}{Isophotal Area} \\
\multicolumn{1}{c}{} &
\multicolumn{1}{c}{} &
\multicolumn{1}{c}{} &
\multicolumn{1}{c}{(mag)} & 
\multicolumn{1}{c}{($10^{41}$ \ergs)} & 
\multicolumn{1}{c}{(kpc$^2$)} 
}

\startdata
\multicolumn{6}{c}{\oiii\ blobs at $z=0.63$}\\
\midrule

\obone\ (OIIIB063s-1) & 02 17 50.244& -05 00 04.159 & $22.04\pm0.14$ & $4.99\pm0.7$ & 1121.2 \\
\obtwo\ (OIIIB063s-2) & 02 19 05.902& -05 13 48.599 & $22.27\pm0.15$ & $4.04\pm0.7$ & 1121.2 \\
\midrule
\multicolumn{6}{c}{\oiii\ blobs at $z=0.83$}\\
\midrule
\obthree\ (OIIIB083m-2) & 02 19 03.734 & -05 11 53.243 & $22.77\pm0.13$ & $4.40\pm0.6$ & 1675.0\\
\obfour\ (OIIIB083m-12) & 02 18 58.701 & -05 12 58.126 & $23.32\pm0.23$ & $2.64\pm0.6$ & 981.9\\
\enddata
\tablenotetext{}{\bf Note:} 
\tablenotetext{a}{The luminosity of the \oiii\ emission lines is derived from the 
magnitude in $NB - BB$ and is not corrected for \textbf{internal} dust attenuation.}
\end{deluxetable*}

\section{Targets for Spectroscopy}\label{sec:target}

We conduct the spectroscopic observations of four \oiii\ blobs at $z\sim0.63$ and $z\sim0.83$. 
Our spectroscopic sample is originally identified as an \oiii\ blob by \cite{yuma17}. 
They conducted the systematic survey of \oii, \oiii, and \ha\ blobs at $z=0.1-1.5$; 
i.e., star-forming galaxies with strong \oii, \oiii, and \ha\ emission lines, respectively, spatially extended over 30 kpc. 
The emission-line extension is measured down to surface flux limit of $1.2\times10^{-18}$ \ergscmarcsec. 
Among 77 blobs at $z=0.1-1.5$ found by \cite{yuma17}, there are 4 and 13 \oiii\ blobs at $z=0.63$ and $z=0.83$, respectively. 
All blobs are crossmatched with the X-ray and radio catalogs 
to primarily check for the AGN existence. 
The X-ray source catalog is obtained from \cite{ueda08}. It is observed with the 
European Photon Imaging Camera \citep{struder01, turner01} 
on the \textit{XMM-Newton} telescope \citep{jansen01}. 
The flux limits of the catalog are $6\times10^{-16}$, $8\times10^{-16}$, $3\times10^{-15}$, 
and $5\times10^{-15}$ \ergscm\ 
in the $0.5-2$, $0.5-4.5$, $2-10$, and $4.5-10$ keV bands, respectively. 
These flux limits approximately correspond to the X-ray luminosities of $10^{42}$ 
\ergs\ at $z\sim0.7$. 
The radio catalog is from \cite{simpson06}. The radio data are observed 
with the Very Large Array (VLA) in 1.4 GHz down to the limit of 100 $\mu$Jy. 
All of the \oiii\ blobs are not detected in X-ray and radio wavelengths. 

The criteria of the spectroscopic targets are that 1) they are among the \oiii\ blobs with 
the largest extension of the \oiii\ emission line 
and/or 2) they are located close to each other so that we can maximize the number of 
the \oiii\ blobs observed within one mask. 
As a result, we select 4 \oiii\ blobs as the spectroscopic targets and observe them with two masks. 
The targets are OIIIB063s-1, OIIIB063s-2, OIIIB083m-2, and OIIIIB083m-12. 
The IDs are from the original catalog by \cite{yuma17}. 
\oiiib\ indicates that it is an \oiii\ blob, a galaxies with the extended \oiii\ emission line. 
Three numbers after \oiiib\ represent the redshifts; for example, 
063 means the blob is at $z=0.63$. 
The letter ``s" or ``m" stands for single or multiple, respectively, 
indicating the number of stellar components in one blob. 
The last numbers show the order of blobs starting from ``1" for the 
blob with the largest extension of the emission line. 
To avoid any confusion, we hereafter refer to OIIIB063s-1, OIIIB063s-2, OIIIB083m-2, and OIIIIB083m-12 
as \obone, \obtwo, \obthree, and \obfour, respectively. 
Figure \ref{fig_spec_target} shows the 2-dimensional (2D) images 
of the \oiii\ emission line, the stellar continuum, and the equivalent-width (EW) map of our 4 targets. 
The emission-line images ($NB-BB$) correspond to the \nbest-\rz\ and \nbnto-\zp\ images 
for the \oiii\ blobs at $z=0.63$ and $z=0.83$, respectively, where \rz\ is defined as 
{\it Rz $\equiv$ (R+2\zp)/3}. 
The contour of the \oiii\ extension is also displayed in the figure. 
The slit direction with the exact width and length is shown in the right panel of Figure \ref{fig_spec_target}. 
Coordinates and photometric properties of these four \oiii\ blobs derived by \cite{yuma17} 
are listed in Table \ref{tab_blob_sample}.

\section{Observation and Data Reduction}\label{sec:data}

\subsection{Subaru/FOCAS}\label{subsec:focas}

We conducted the optical spectroscopic follow-up observations 
with Subaru/Faint Object Camera and Spectrograph \citep[FOCAS; ][]{kashikawa02} 
on October $21-22$, 2014 and December $2-4$, 2015 (S14B-130 and S15B-059; PI: S. Yuma). 
Details about the Subaru/FOCAS spectroscopic observations
are described in \cite{yuma17}. 
Briefly speaking, we performed the observations in the multi-object spectroscopy 
(MOS) mode with the VPH450 grating and the VPH850 grating with the SO58 order-cut filter. 
The gratings cover $3800-5250$ \AA\ with a dispersion of 0.37 \AA/pixel 
and $5800-10500$ \AA\ with 1.17 \AA/pixel, respectively. 
We adopted the slit width of $0\farcs8$ providing the spectral resolutions of 1700 and 750 
for VPH450 and VPH850+SO58, respectively. 
We used 2 MOS masks to observed 4 \oiii\ blobs. 
Mask1 contains only \obone\ which is the largest \oiii\ blob at $z=0.63$, while the other 3 blobs 
(\obtwo, \obthree, and \obfour) are so close to each other that we can place them all in Mask2. 
The on-source exposure times for Mask1 are 180 minutes and 120 minutes in VPH450 and VPH850+SO58, respectively. 
For Mask2, the exposure times are 260 minutes and 100 minutes. 
The sky when we observed the target in Mask1 was quite clear in that we obtain the seeing size of $\sim0\farcs6-0\farcs7$. 
Unfortunately, the weather was getting worse when we observed Mask2. The seeing size is roughly $\sim0\farcs9$. 
We carefully designed the slit direction to cover the longest extent of the \oiii\ emission line 
as shown with the magenta color in Figure \ref{fig_spec_target}. 
Note that, in Mask2, we intended to do so with \obthree, which is the second largest \oiii\ blob at $z=0.83$. 
Because all slits in one mask are required to arrange in the identical direction, the slit directions for \obtwo\ and \obfour\ are consequently not oriented to the longest extension of their \oiii\ emission lines.

\begin{deluxetable}{lcc}[!h]
\tabletypesize{\small}
\tablewidth{0pt}
\tablecolumns{3}
\tablecaption{Details of optical and near-infrared spectroscopic observations for 4 \oiii\ blobs at $z=0.63$ and $z=0.83$  \label{tab_focas_obs}}
\tablewidth{0pt}
\tablehead{
\multicolumn{1}{c}{Ins./Grating+Filter} &
\multicolumn{1}{c}{Exposure time} &
\multicolumn{1}{c}{$3\sigma$ flux limit}\\
\multicolumn{1}{c}{} &
\multicolumn{1}{c}{(minutes)} &
\multicolumn{1}{c}{(\ergscm)}
}
\startdata
\multicolumn{3}{l}{Mask1: \obone}\\
\midrule
FOCAS/VPH450 & 180 & $\simeq3.8\times10^{-17}$\\
FOCAS/VPH850+SO58 & 120 & $\simeq1.1\times10^{-18}$\\
MOSFIRE/Y & 285 & $\simeq3.1\times10^{-18}$ \\
\midrule
\multicolumn{3}{l}{Mask2: \obtwo, \obthree, and \obfour}\\
\midrule
VPH450 & 260 & $\simeq1.1\times10^{-17}$ \\
VPH850+SO58\tablenotemark{a} & 100 & $\simeq1.0\times10^{-18}$
\enddata
\tablenotetext{}{\bf Note:} 
\tablenotetext{a}{For \obtwo, we discard two observed frames to avoid the cosmic ray at the wavelength of the \hb$\lambda4861$ emission line. Therefore, the $3\sigma$ flux limit of \obtwo\ is $\simeq3.7\times10^{-18}$ \ergscm.}
\end{deluxetable}

Data reduction has been carried out with the FOCASRED package,  
which is the specific pipeline on Imaging Reduction and Analysis Facility (IRAF) for Subaru/FOCAS data reduction. 
We started with bias subtraction, flat fielding, and distortion correction. 
Wavelength calibration is then performed by using the ThAr lamp and 
OH airglow emission lines for VPH450 and VPH850+SO58 spectra, respectively. 
After we subtracted the sky background, we stacked the spectra 
and extracted them along the spatial direction to create one-dimensional (1D) spectra. 
The extracting width is designed to cover the entire spatial extension of the emission line. 
Flux calibration is carried out by using the standard star G191-B2B observed with 
the same slit width. 
It is extracted to 1D spectrum with the same extraction width as the target. 
The slit loss is automatically corrected during the flux calibration process. 
The $3\sigma$ flux limits are summarized in Table \ref{tab_focas_obs}. 

\begin{figure*}
	\centering
	\includegraphics[width=0.95\textwidth]{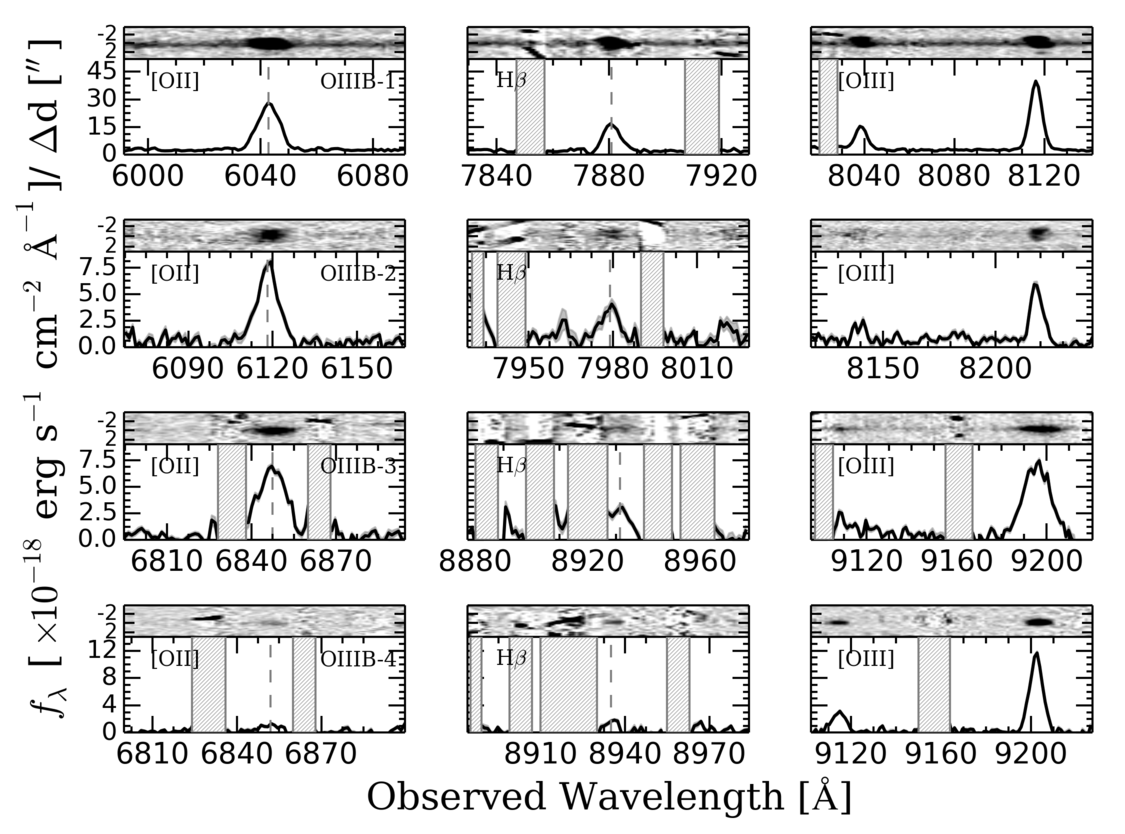}
	\caption{Subaru/FOCAS spectra of \obone, \obtwo, \obthree, and \obfour\ from the top to the bottom panels, respectively. 
The left, middle, and right columns show the \oii, \hb, and \oiii\ emission lines of each object, respectively. 
In each plot, we illustrate the 2D spectrum in the top panel centering at the center of the stellar component of the blob, while the bottom panel shows 1D spectrum at the corresponding observed wavelengths. 
The centers of \oii\ and \hb\ lines are indicated with the dashed gray lines. 
The shade regions in some figures indicate the area with poor S/N ratios in the spectrum . 
	}\label{fig_spec_lines}
\end{figure*}

\subsection{Keck/MOSFIRE}

Near-infrared (NIR) observation was conducted only for \obone\ 
with Multi-Object Spectrometer For Infrared Exploration \citep[MOSFIRE;][]{mclean12} 
attached on Keck I telescope. 
\obone\ was observed as a filler target 
with the Y-band filter on January $3-4$, 2014 (S15B-075; PI: M. Ouchi). 
The main objective of the MOSFIRE proposal is to confirm the Lyman alpha emitters at $z\sim5.7$. 
The spectral resolution is $3388$ with the standard $0\farcs7$ slit width. 
The total exposure time is 4.45 hours with seeing size of $\sim0\farcs7$. 
Data reduction is performed with MOSFIRE data reduction pipeline\footnote{https://keck-datareductionpipelines.github.io/MosfireDRP/} 
following the standard processes for reducing the spectra and flux calibration 
with the standard stars taken during the observations. 
The exposure time and the limiting flux are also listed in Table \ref{tab_focas_obs}.

\section{Results}\label{sec:results}

\subsection{Systemic redshifts}

Before we investigate the spectra in details, we first check whether our \oiii\ blob sample is at the targeted redshifts. 
We examine the emission lines detected in the VPH850+SO58 spectra. 
The \oiii\ blobs at $z\sim0.63$ and $z\sim0.83$ are originally identified 
by the narrowband technique using the {\it NB816} and {\it NB921} images, respectively. 
So we expect the \oiii\ emission line to be around 
$8090-8210$ \AA\ for \obone\ and \obtwo\ and 
$9130-9262$ \AA\ for \obthree\ and \obfour. 
In fact, \cite{yuma17} already confirmed that 
the emission line detected in the wavelength ranges mentioned above is 
\oiii$\lambda5007$ based on the spectroscopic confirmation of 
\oii$\lambda3727$, \hb$\lambda4861$, 
and \oiii$\lambda4959$ emission lines at the corresponding wavelengths. 
They confirmed that \obone\ and \obtwo\ are at 
$z=0.621$ and at $z=0.641$, respectively, whereas 
\obthree\ and \obfour\ are both at $z=0.838$. 
We  fit the \oii, \hb, and \oiii\ emission lines with a single Gaussian profile 
and adopt the center of the Gaussian profile that is best fitted to the \oiii$\lambda5007$ emission line 
as the spectroscopic redshift for each \oiii\ blob (Table \ref{tab_spec_results}). 
Our resulting spectroscopic redshifts are consistent with those derived by \cite{yuma17}. 


\subsection{Spectroscopic Properties of Emission lines}\label{subsec:spec}

The Subaru/FOCAS spectra of 4 \oiii\ blobs at $z\sim0.63-0.83$  are illustrated in Figure \ref{fig_spec_lines} 
at the observed wavelengths of the \oii$\lambda3727$, \hb$\lambda4861$, and 
\oiii$\lambda\lambda4959,5007$ emission lines. 
In each plot, we show the two-dimensional spectral image in the top panel and the one-dimensional spectrum 
in the bottom one. 
All the lines for a given blob are plotted with the same flux scale to provide the idea of the strength of the emission lines 
as compared to one another. 
The top row of Figure \ref{fig_spec_lines} shows that 
the stellar continuum is significantly detected in \obone. 
The excessive extension of the emission lines beyond the stellar components is clearly seen 
in the 2D spectral image. 

Although we marginally detect the stellar continuum of \obtwo, 
the emission lines of \obtwo\ seem to be extended over the stellar continuum in the spatial direction. 
Likewise, we are able to see the spatial extension of the \oiii$\lambda5007$ emission line of \obthree. 
\obthree\ shows the significantly larger spectral line width of the \oiii$\lambda5007$ emission as compared with \obone\ and \obtwo. 
The full width at half maximum (FWHM) of the \oiii$\lambda5007$ line of \obthree\ is $419.5\pm 76.8$ \kms\ after correcting for the instrument, while it is in the order of $150-200$ \kms\ for \obone\ and \obtwo. 
For \obfour, the stellar continuum is unfortunately not detected, but we can detect the \oiii\ emission line with the high signal-to-noise (S/N) ratio. 
However, the \oii\ and \hb\ emission lines are marginally detected with the S/N ratios of $2-3\sigma$. 
The observed flux of each emission line is derived by fitting the 1D spectrum with a single Gaussian profile and corrected for the dust attenuation, which is explained in Appendix \ref{appen:dust}.  
In short, we derive the dust attenuation of \obone\ and \obtwo\ from the Balmer decrement using the \hg/\hb\ line ratios. 
We find that the color excesses $(E(B-V))$ estimated from the Balmer decrement is in agreement 
with those derived by fitting the observed spectral energy distribution of the blob with the stellar population synthesis models (SED fitting method; Appendix \ref{appen:dust}). 
Therefore, for \obthree\ and \obfour\ whose \hg\ emission lines are not detected, we adopt the color excesses derived by the SED fitting method.  
The nebular color excesses of \obthree\ and \obfour\ are then calculated by using the relation between the nebula and the stellar dust extinctions in \cite{calzetti00}. 
More details on the SED fitting procedures of the \oiii\ blobs are explained in Section \ref{subsec:discovery} and \cite{yuma17}. 
We summarize the dust-corrected fluxes of the \oii$\lambda3727$, \hb$\lambda4861$, and \oiii$\lambda5007$ emission lines with corresponding $1\sigma$ uncertainties in Table \ref{tab_spec_results}. 
The extended features seen in \obone, \obtwo, and \obthree\ are just confirmed that the blob selection method by \cite{yuma17} is efficient in selecting galaxies with the spatially extended emission line.

\begin{figure*}
	\centering
	\includegraphics[width=0.9\textwidth]{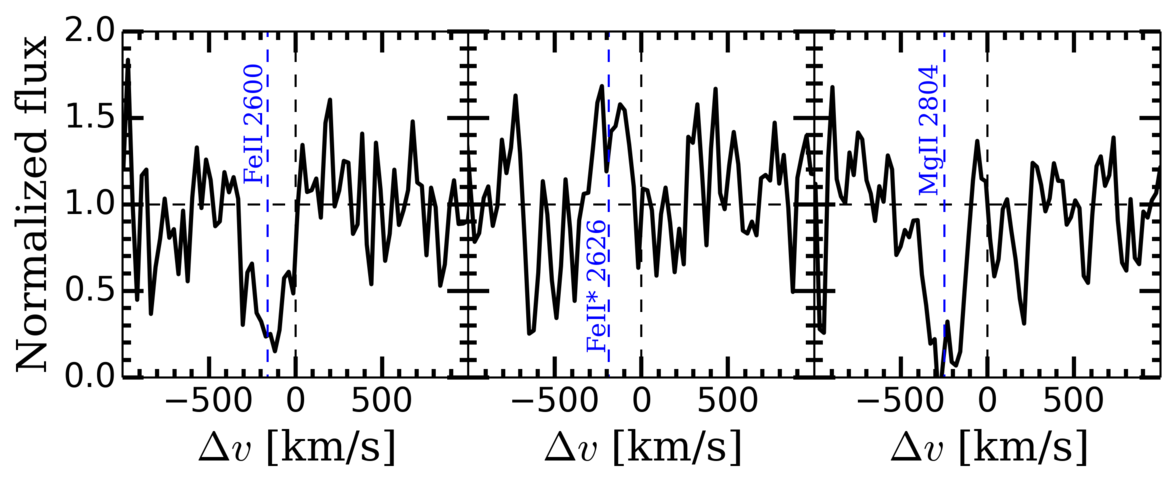}
	\caption{The blueshifted \feii$\lambda2600$ absorption ({\it left}), \feiis$\lambda2626$ 
fine-structure emission ({\it middle}), and \mgii$\lambda2804$ absorption lines ({\it right}) of \obone. 
The observed flux is normalized to the continuum spectrum. The blue vertical dashed lines indicate 
the centroids of velocity offsets of each line.
	}\label{fig_outflow}
\end{figure*}

\begin{deluxetable}{l rrrr}[!h]
\tabletypesize{\small}
\tablewidth{0pt}
\tablecolumns{5}
\tablecaption{Spectroscopic redshifts and dust-corrected fluxes of \oii, \hb, and \oiii\ emission lines for 4 \oiii\ blobs at $z=0.63-0.83$\label{tab_spec_results}}
\tablewidth{0pt}
\tablehead{
\multicolumn{1}{c}{Target} &
\multicolumn{1}{c}{$z_{spec}$} &
\multicolumn{3}{c}{Observed Flux ($\times10^{-16}$ \ergscm)} \\
\multicolumn{1}{c}{} & 
\multicolumn{1}{c}{} & 
\multicolumn{1}{c}{\oii} & 
\multicolumn{1}{c}{\hb} &
\multicolumn{1}{c}{\oiii} 
}
\startdata
  \obone 	 & 0.6210	 & $16.67\pm0.29$ & $5.02\pm0.27$ & $12.16\pm0.25$ \\
  \obtwo 	 & 0.6413 & $7.33\pm0.60$ & $1.53\pm0.31$ & $3.58\pm0.65$ \\
  \obthree 	& 0.8365	& $6.12\pm0.51$ & $0.71\pm0.35$ & $5.48\pm0.68$ \\ 
  \obfour  	& 0.8379	& $0.31\pm0.11$ & $0.30\pm0.12$ & $2.02\pm0.12$ 
\enddata
\end{deluxetable}

\subsection{Outflow Signature}\label{subsec:outflow}

The galaxies with spatially extended emission lines like \oiii\ blobs are 
thought to be in an ongoing process of the large-scale outflow. 
The gas outflow is already confirmed in \oii\ blobs at $z\sim1.2$ 
with Subaru/FOCAS, VLT/VIMOS, and Magellan/LDSS spectra of the \oii\ blobs 
at $z\sim1.2$ \citep{yuma13, harikane14}; however, 
this is the first time to confirm if the \oiii\ blobs are in the middle of the outflow process. 
The blueshifted interstellar absorption lines are one of the most common methods 
that are used to study the outflow process of the galaxies. 
Unfortunately, the spectra of three out of four blobs in the VPH450 grating, 
in which we expect to detect the absorption lines, have too low S/N ratios. 
We only detect the absorption lines in the \obone\ spectrum. 
Figure \ref{fig_outflow} shows the blueshifted 
\feii$\lambda2600$ absorption, \feiis$\lambda 2626$ fine-structure emission, 
and \mgii$\lambda2804$ absorption lines. 
They are all blueshifted from the systemic redshift, indicating the gas outflow from the galaxy. 
We fit the blueshifted absorption and emission lines with a single Gaussian profile. 
The velocity offsets of the \feii, \feiis, and \mgii\ lines are $-160$ \kms, 
$-200$ \kms, and $-270$ \kms, respectively. 
It is important to note that the \mgii\ absorption line 
is possibly affected by strong resonant emission at the systemic velocity. 
This infilling emission can shift the centroid of the 
\mgii\ velocity offset in the order of tens \kms\ \citep[e.g., ][]{prochaska11}. 
\cite{erb12} studied the outflow of normal star-forming galaxies at $z=1-2$ by using 
various interstellar absorption lines including those detected in this work. 
They found that the velocity offsets of the \feii$\lambda2600$ and \mgii$\lambda2804$ 
absorption lines 
of star-forming galaxies at $z=1-2$ with the stellar masses and star formation rates comparable to those of \obone\ range between  $-190$ \kms\ and $-130$ \kms. 
The \feiis\ and \mgii\ lines of \obone\ show the velocity offsets slightly higher than those of the normal star-forming galaxies.

\begin{figure*}
	\centering
	\begin{tabular}{cc}
		\includegraphics[width=0.45\textwidth]{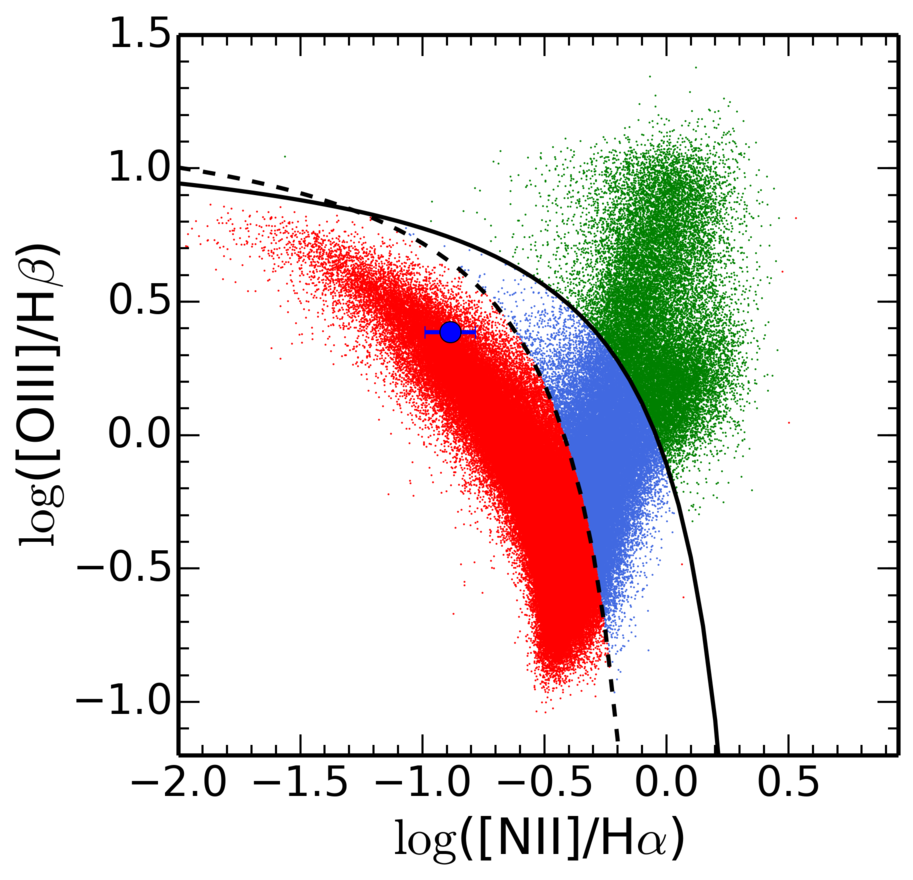} &
		\includegraphics[width=0.475\textwidth]{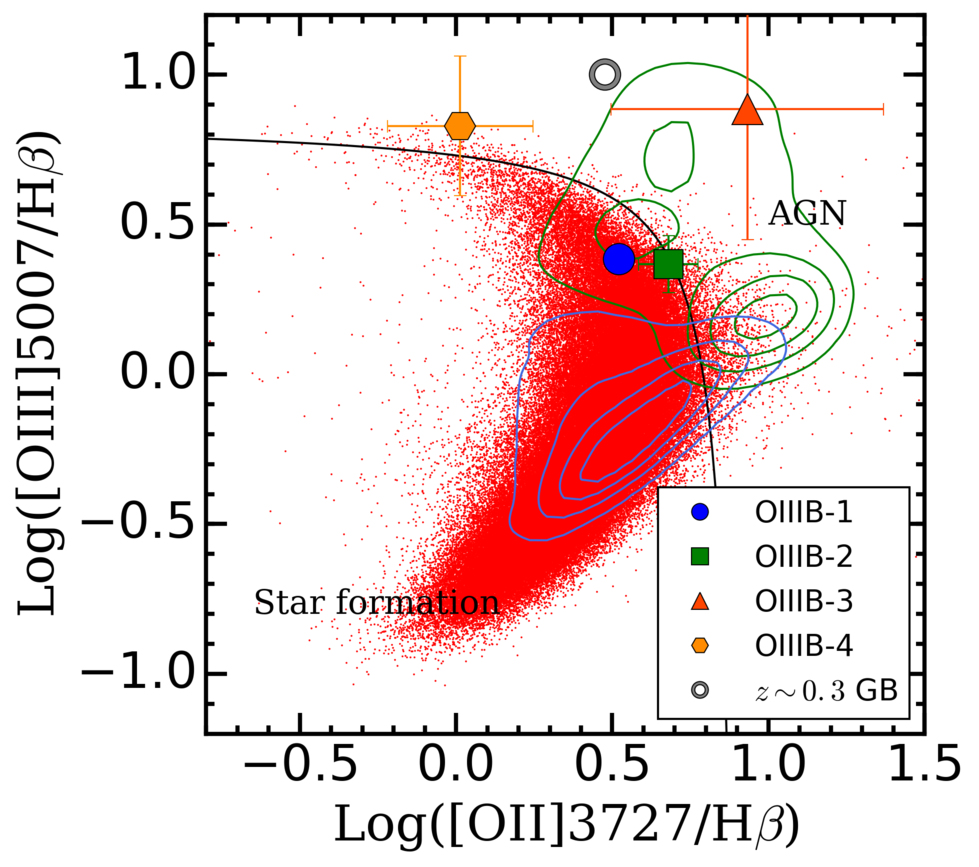}
	\end{tabular}
	\caption{{\it Left panel:} A BPT diagram showing the \oiii/\hb\ ratio against the \nii/\ha\ ratio. 
	The blue circle indicates \obone, while local galaxies taken from SDSS DR7 data are shown 
	in three different colors. 
	The red, blue, and green dots are for star-forming galaxies, composites, and AGNs, respectively. 
	The black dashed and solid curves are the criteria for separating pure star-forming galaxies, composites, and AGNs by \cite{kauffmann03} and \cite{kewley01}, respectively. 
	{\it Right panel:} \oiii/\hb\ versus \oii/\hb\ or the blue diagram of four \oiii\ blobs at $z=0.63$ and $z=0.83$.  
	The line ratios are already corrected for dust attenuation. 
	The blue circle is \obone\ and the green square is \obtwo\ at $z\sim0.63$. 
	\obthree\ and \obfour\ at $z\sim0.83$ are indicated with the red triangle and 
	the orange hexagon, respectively. 
	Red dots represent the local star-forming galaxies classified by the BPT diagram in the left panel. 
	Blue and green contours show the distribution of composites and AGNs, respectively. 
	The open gray circle shows a Seyfert type-2 AGN (J2240-0927) at $z\sim0.3$ 
	with spatially extended \oiii\ emission line, which is called a green bean \citep{schirmer13}. 
	The black solid line divides the blue diagram into two regions of star formation and AGN \citep{lamareille10}. 
	}\label{fig_agn}
\end{figure*}

\subsection{AGN Diagnostics}\label{subsec:agn}

An AGN is one of plausible energy sources 
providing hard ionizing photons that can cause the emission lines at large scale. 
\cite{yuma17} carried out a primary check for the AGN contribution in the \oiii\ blobs 
by cross-matching the X-ray and radio catalogs obtained by \cite{ueda08} and 
\cite{simpson12}, respectively. 
They found no X-ray or radio counterpart for any \oiii\ blobs in our sample 
down to the X-ray luminosity of $10^{42}$ \ergs\ 
as mentioned in Section \ref{sec:target}. 
This luminosity limit is $\sim0.5$ dex brighter than the faintest bin of 
the x-ray luminosity functions (XLFs) 
of AGNs at similar redshifts \citep[e.g., ][]{fotopoulou16, ranalli16}. 
So we know that the \oiii\ blobs are not an unobscured AGN, but 
the possibility that the \oiii\ blobs are faint or heavily obscured AGNs 
cannot yet be ruled out. 

To further investigate AGN signature in the targets, 
we first plot the Baldwin, Phillips, and Terlevich (BPT) digram in the left panel of Figure \ref{fig_agn}. 
The BPT diagram is a plot of the \oiii/\hb\ ratio versus the \nii/\ha\ ratio \citep{baldwin81}. 
\cite{kewley06} used this diagram to efficiently classify their sample into 
the star-forming galaxies, composites, and AGNs. 
Compared with star-forming galaxies, AGNs have the higher ionization state resulting in higher values of both \oiii/\hb\ and \nii/\ha\ ratios. 
We plot local galaxies ($z=0.04-0.10$) obtained from the Sloan Digital Sky Survey (SDSS) Data Release 7 
\citep[DR7;][] {abazajian09} in the left panel of the figure and separate 
them into pure star-forming galaxies (red), composites (blue), and AGNs (green) 
based on their locations on the BPT diagram. 
\obone\ is the only object that has the near infrared spectrum available. 
It is clearly seen from the figure that \obone\ lies exactly 
on the distribution of the local star-forming galaxies indicating that it is a normal star-forming galaxy. 

For the remaining three \oiii\ blobs, we use the blue diagram, which is a plot 
between the dust corrected \oiii/\hb\ and \oii/\hb\ emission line ratios 
(the right panel of Figure \ref{fig_agn}). 
AGNs can be distinguished from the star-forming galaxies with the solid curve shown in the right panel of Figure \ref{fig_agn} \citep[e.g.,][]{lamareille10, harikane14}. 
On the blue diagram, we display the local SDSS galaxies that we already categorized into star-forming galaxies, composites, and AGNs based on the BPT diagram as a reference. 
The local star-forming galaxies and composites are largely overlapped with each other 
in the region below the curve that is used to separate the star formation activity from AGNs. 
On the other hand, AGNs can be distinguished from the star formation galaxies and composites efficiently. 
\cite{lamareille10} stated that the contamination of AGN in the region of star-forming galaxies on the blue diagram is roughly 16\%. 
We also plot a Seyfert type-2 AGN (J2240-0927) at $z=0.326$ that  
shows the extended \oiii\ emission line \citep{schirmer13}. We discuss the similarity and difference 
between our \oiii\ blobs and this AGN in Section \ref{subsec:discovery}.

From the blue digram in the right panel of Figure \ref{fig_agn}, 
\obone\ is still confirmed to be in the pure star formation region consistent with 
its location on the BPT diagram. 
\obtwo, on the other hand, shows the higher \oii/\hb\ ratio and get closer to the separation curve. 
Thus we cannot rule out the possibility that \obtwo\ might be fueled by an AGN activity. 
The other two \oiii\ blobs at $z\sim0.83$ show the remarkably high \oiii/\hb\ ratios. 
\obthree\ is clearly located in the AGN region. 
This is in good agreement with its emission-line width mentioned in Section \ref{subsec:spec}. 
The \oiii$\lambda5007$ line width of 419.5 \kms 
suggests that \obthree\ is plausibly the type-2 Seyfert, whose typical line widths are roughly $\sim500$ \kms. 

\obfour\ shows the remarkably high \oiii/\hb\ ratio but low \oii/\hb\ ratio 
(the right panel of Figure \ref{fig_agn}). 
The high ratio of \oiii\ to \hb\ emission lines places \obfour\ above the separated curve in the blue diagram. 
As seen in the blue diagram, some fraction of the local star-forming galaxies distribute 
in the same region as \obfour, while AGNs do not. 
It is likely that \obfour\ is one of the star-forming galaxies with the strongest \oiii/\hb\ ratio. 
In conclusion, with the emission-line diagnostics, we can confirm that \obone\ and \obfour\ are star-forming galaxies, 
whereas \obthree\ is an obscured AGN. 
However, there is no clear conclusion for \obtwo; 
it can be a star-forming galaxy, composite galaxy, or AGN.

\begin{figure*}
	\centering
	\begin{tabular}{cc}
		\includegraphics[width=0.45\textwidth]{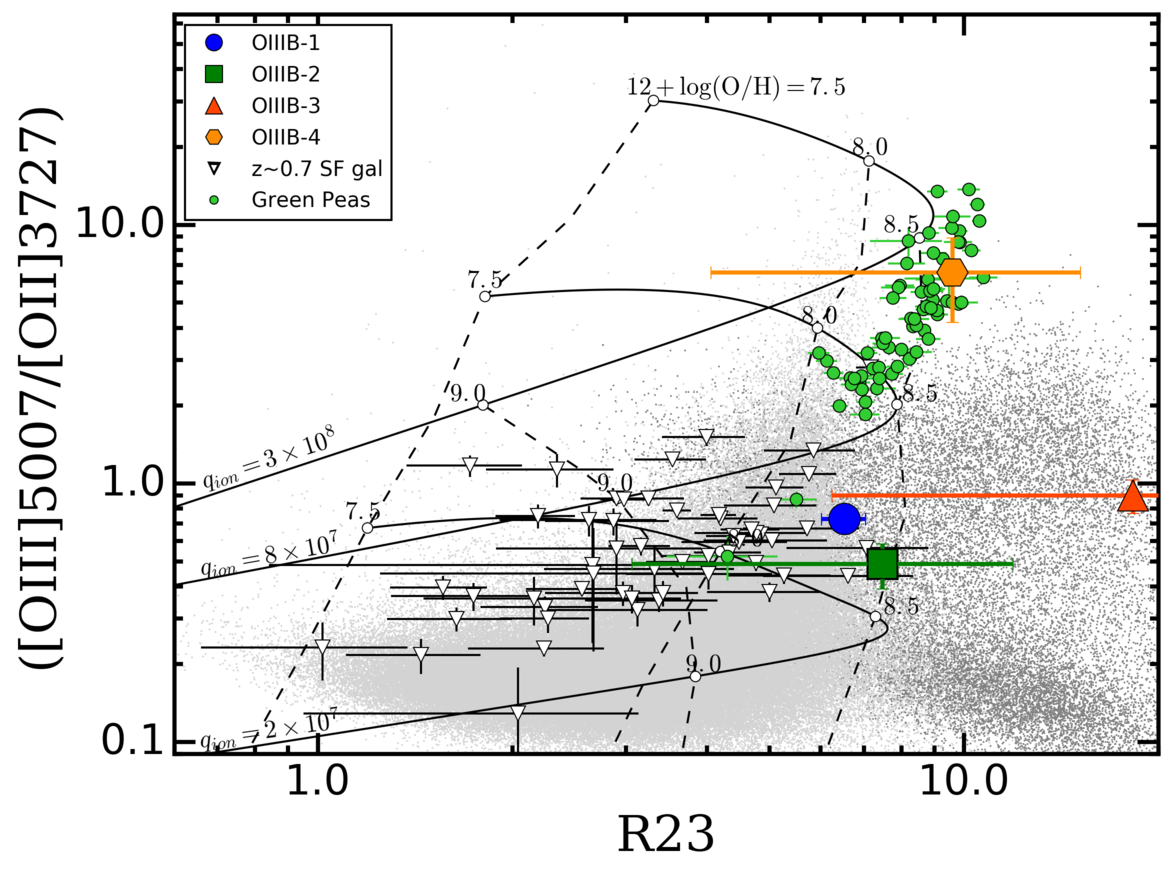} & 
		\includegraphics[width=0.45\textwidth]{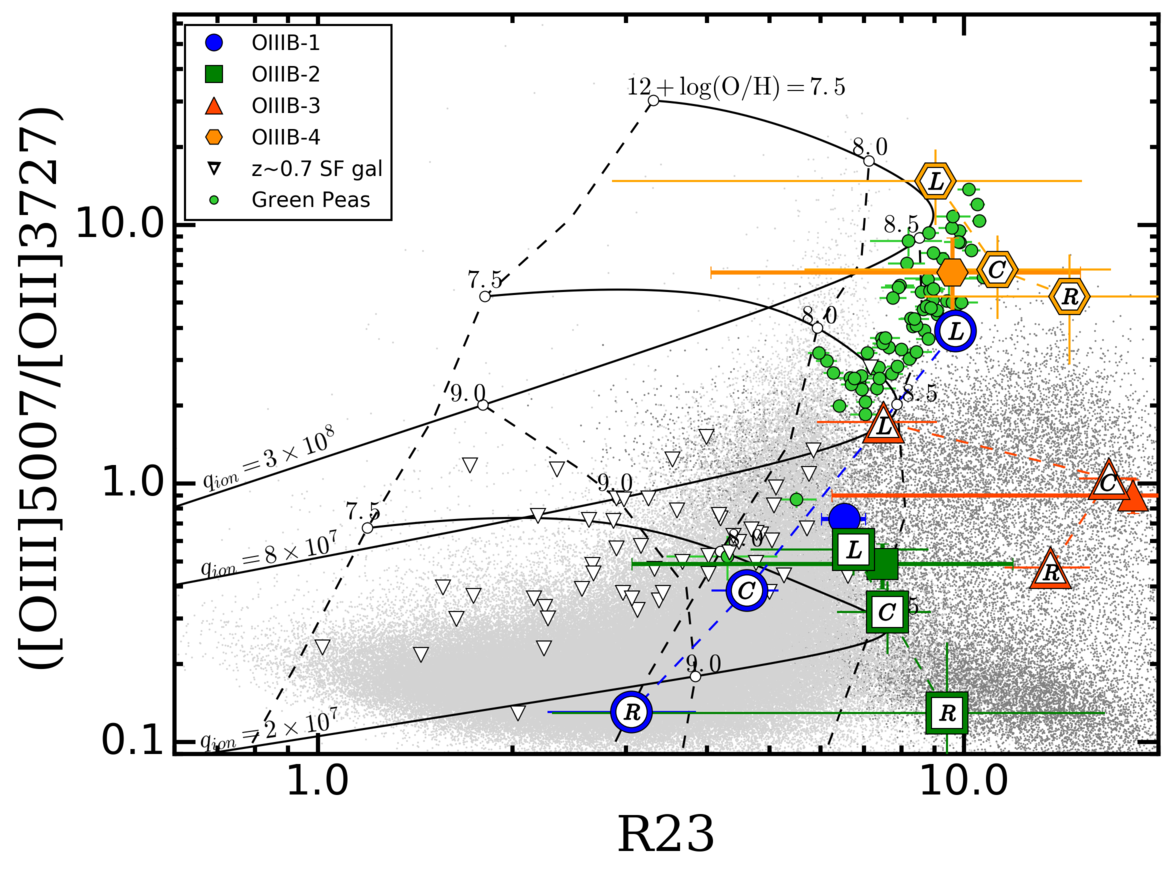} \\
	\end{tabular}
	\caption{The relation between \oiii/\oii\ ratio and {\it R23} index. 
	{\it Left panel}: Colored symbols represent \oiiib s at $z\sim0.7$ as in previous figures. 
	Gray and dark gray dots are the SDSS star-forming galaxies and AGNs, respectively. 
	Open triangles represent star-forming galaxies at $z=0.5-0.9$ by \cite{lilly03}. 
	Green circles are for the local green peas with the strong \oiii\ emission line by \cite{cardamone09}.
	Each solid black line shows a track of photoionization models at a constant  
	ionization parameters of $q_{ion}= 2\times10^7, 8\times10^7$, and $3\times10^8 {\rm cm\,s}^{-1}$ 
	with 12+$\log$(O/H) in the ranges of $7.5-9.0$ \citep{kewley02}. 
	The dashed lines connect the photoionization model tracks at a constant 12+$\log$(O/H).
	{\it Right panel}: \oiii/\oii\ ratio vs {\it R23} index of the radial profiles of the \oiii\ blobs. 
	Opened symbols with the letters ``L", ``C", and ``R" refer to the left, central, and right parts of the blobs in Figure \ref{fig_o32profile} (More details in context). 
	Other symbols and lines are identical to those in the left panel. 
	}\label{fig_o32r23}
\end{figure*}

\subsection{Properties of ISM}\label{subsec:ism}

\cite{nakajima13} and \cite{nakajima14} show that the diagram between the \oiii/\oii\ line ratio 
and the ${\it R23}$ index 
is efficient to investigate an ionization parameter and metallicity of a star-forming galaxy
simultaneously. 
The emission-line ratio of an atom at different ionization states such as the \oiii/\oii\ ratio 
strongly depends on the ionization parameter of gas, $q_{ion}$, which 
is the ratio of the ionizing photon flux per unit area 
to the number density of hydrogen atoms. 
The ionization parameter $q_{ion}$ corresponds to the dimensionless ionization parameter $U$ as $U\equiv q_{ion}/c$, where $c$ is the speed of light \citep{kewley02}. 
The {\it R23} index is defined as 
\begin{equation}
	R23 = \frac{[{\rm O}\,\textsc{II}]\lambda3727 + [{\rm O}\,\textsc{III}]\lambda\lambda4959,5007}{\rm H\beta}. 
\end{equation}
The {\it R23} index is one of the good indicators for estimating metallicities of the galaxies 
\citep[e.g.,][]{pagel79, kewley02}, although it slightly depends on the ionization parameter.

We plot the \oiii/\oii\ ratios of the \oiii\ blobs
against their {\it R23} indices in Figure \ref{fig_o32r23} to investigate 
the properties of their interstellar media (ISM). 
The local SDSS star-forming galaxies (gray dots), local SDSS AGNs (dark gray dots), 
and star-forming galaxies at $0.47< z <0.92$ 
\citep[open triangles; ][]{lilly03} are also plotted as a reference. 
Another population that we plot in the Figure as a reference is Green Peas (hereafter GPs). 
GPs are the compact star-forming galaxies at $z=0.112-0.360$ with 
the strong \oiii\ emission line 
with the rest-frame equivalent width up to 1,000 \AA\ \citep{cardamone09}. 
The \oiii\ emission line of GPs falling into the $r$ band makes this type of galaxies looks green 
in the $g,r,i$ composite image. 
In addition to the observed galaxies at various redshifts, 
we plot photoionization models at constant ionization parameters 
of $q_{ion}=2\times 10^7, 8\times10^7,$ and $3\times10^8$ \cms\ with varying metallicities. 
The models with the constant metallicities are linked with dashed lines and 
the values of the metallicities in terms of $12+\log ({\rm O/H})$ 
are indicated by the open circles. 
The local SDSS star-forming galaxies 
show a variety of $R23$ indices ranging from $1.0$ to $10.0$, while 
their \oiii/\oii\ ratios are mostly below $1.0$. 
Likewise, star-forming galaxies at $z\sim0.7$ are distributed in the same area on the plot. 
It is indicated that most of the star-forming galaxies at $z=0$ and $z\sim0.7$ 
have ionization parameters between $q_{ion}=2\times10^7$ \cms\ and $q_{ion}=8\times10^7$ \cms\ 
but have wide ranges of metallicities. 
On the other hand, the star-forming GPs 
with the strong \oiii\ emission line show the high \oiii/\oii\ ratios 
and the $R23$ indices. 
According to the figure, the GPs at $z=0.1-0.4$ show the ionization parameters 
higher than $8\times10^7$ \cms, 
which is clearly higher than the normal star-forming galaxies at $z=0$ and $z=0.7$. 
Due to the strong \oii\ and \oiii\ emission, the SDSS AGNs at $z\sim0$ show high $R23$ indices 
that is relatively greater than those of the star-forming galaxies at $z=0$ and $z\sim0.7$. 
Meanwhile, the \oiii/\oii\ line ratios of the local AGNs 
are distributed in the same range as those of the star-forming galaxies.

\obone\ and \obtwo\ are located in the same region as the local and $z=0.7$ star-forming galaxies 
suggesting similar ionization parameters and metallicities. 
\obone\ and \obtwo\ should have 
the ionization parameter slightly less than $q_{ion}=8\times10^7$ \cms. 
Note that the inferred ionization parameter and metallicies of \obtwo\ only apply if it is 
a star-forming galaxy. 
\obthree\ show huge error bars in the $R23$ index due to the low S/N ratio of the \hb\ emission line. 
As \obthree\ is classified as an AGN in Section \ref{subsec:agn}, 
it is obvious that its location is consistent with the AGNs found in the local universe. 
In contrast, \obfour\ is located in the totally different region in Figure \ref{fig_o32r23}. 
\obfour\ show the impressively high \oiii/\oii\ ratio indicating that the ionization parameter 
$q_{ion}$ should be roughly $3\times10^8$ \cms. 
This high \oiii/\oii\ ratio of \obfour\ agrees well with those of the GPs at $z=0.1-0.4$.

\begin{figure}
	\centering
	\includegraphics[width=0.45\textwidth]{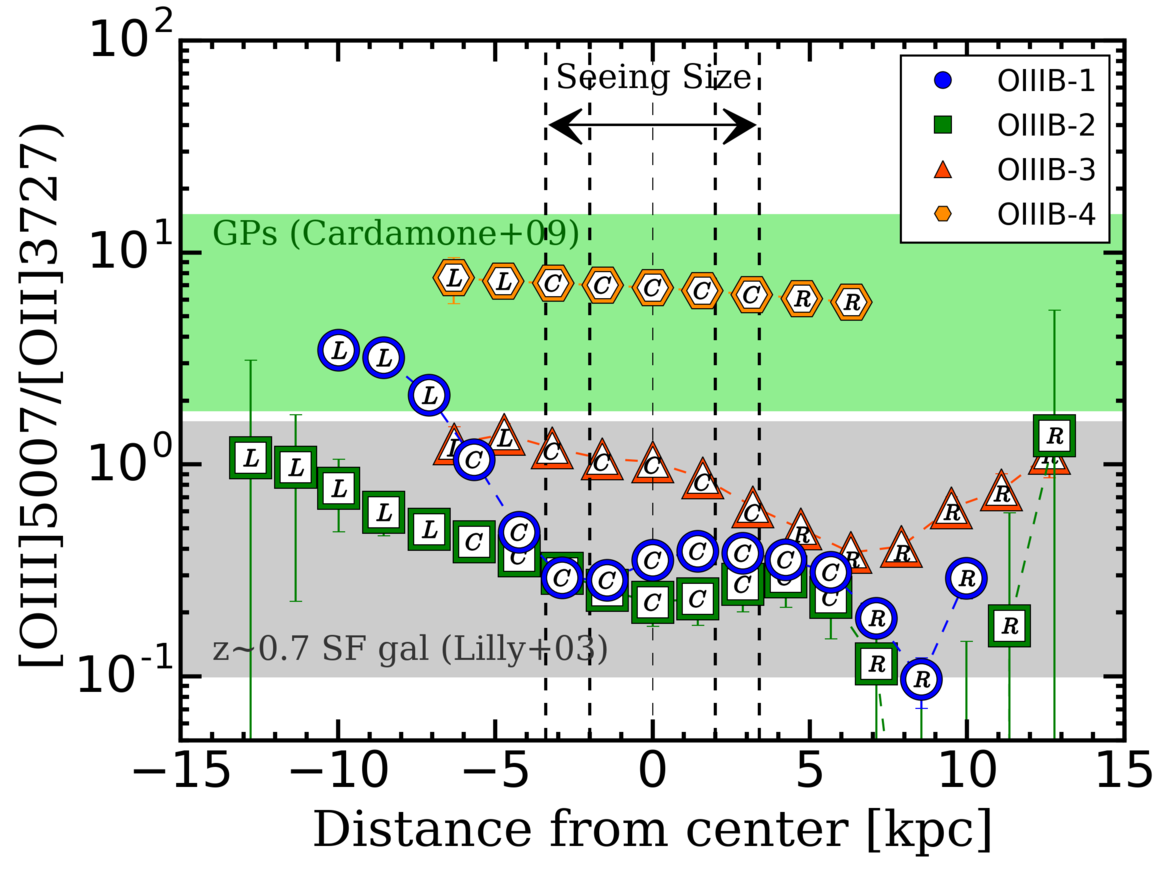}
	\caption{
	Radial profiles of \oiii/\oii\ ratios of 4 \oiiib s at $z\sim0.7$ as a function of the distance from the center of the blobs. 
	The left-to-right radial profile is drawn from the top to the bottom part of the 2D spectrum in Figure \ref{fig_spec_lines}. 
	The symbols of 4 \oiii\ blobs are identical to the previous figures. 
	The letters "L", "C", and "R" indicate the regions that we use to determine 
	the left, central, and right components of a given blob. 
	The grey area represents the typical line ratios of star-forming galaxies at $0.47< z<0.92$ \citep{lilly03}. 
	The green region indicates the line ratios of the green peas at $z=0.112-0.360$ by \cite{cardamone09}. 
	The average seeing size of the data is also shown in the figure. 
	}
	\label{fig_o32profile}
\end{figure}

\subsection{Radial Profile of the \oiii\ blobs}\label{subsec:profile}

We further investigate the nature of the \oiii\ blobs by plotting their \oiii/\oii\ line ratios 
as a function of a distance from the center of the blobs. 
By observing the \oiii\ blobs with the slits finely aligned  
along the most extended direction of the \oiii\ emission line (see Figure \ref{fig_spec_target})\footnote{This is not the case for \obtwo\ and \obfour, whose slit direction is fixed to that of \obthree, the brightest \oiii\ blob in the same mask (see Section \ref{subsec:focas}).}, 
we are able to examine the spectroscopic properties of the outer region of the \oiii\ blobs. 
The radial profile of the emission line is constructed from the background-subtracted 2D spectrum 
by integrating the emission line at each spatial position  
along the wavelength direction covering the full wavelength range of the best-fitted Gaussian profile. 
This wavelength range covers all spatial components of the emission line. 
The stellar continuum profiles at the wavelengths blueward and redward 
of each emission line for a given blob are estimated by creating the continuum profiles 
using the identical wavelength range as used to create the emission-line profile. 
Then we interpolate both continuum profiles to obtain the average continuum profile at 
the wavelength of the emission line. 
In this way, we can ensure that the emission-line profile is not over- or under-estimated. 
The spatial profile of the \oiii/\oii\ ratio 
is obtained after carefully removing the estimated stellar continuum 
and shown in Figure \ref{fig_o32profile}. 
The negative values of the distance from the center refer to the component at the top of 2D spectral image in Figure \ref{fig_spec_lines}. 
For comparison purpose, we show the shaded regions where the GPs at $z=0.1-0.4$ and star-forming galaxies at $z\sim0.7$ are typically located. 

It is seen from Figure \ref{fig_o32profile} 
that all of our four samples extend beyond the seeing size of the spectral images. 
The center of \obone, \obtwo, and \obthree\ lie in the region where the typical star-forming galaxies at $z\sim0.7$ are located. 
Interestingly, the left part of the \obone\ profile at the distance larger than $5$ kpc show the high \oiii/\oii\ ratio consistent with the GPs at $z-0.1-0.4$. 
Similarly, the \oiii/\oii\ ratio of \obtwo\ increases as a function of the distance from the center on both sides, albeit the large uncertainty. 
\obthree\ shows the \oiii/\oii\ ratio of roughly $1-2$, which is slightly higher than the central parts of \obone\ and \obtwo, but it is still in the ranges of the typical star-forming galaxies.  
Finally, the \oiii/\oii\ radial profile of \obfour\ 
seems to be almost constant at roughly \oiii/\oii $=5-10$ over $\sim14$ kpc. 
It is noteworthy that the spatial extension of \obfour\ is smaller than the other three blobs 
because the slit direction is unfortunately not designed to observe the longest axis of \obfour. 
The \oiii/\oii\ ratios are significantly greater than those of the typical star-forming galaxies 
and consistent with those of the compact GPs at $z-0.1-0.4$. 
This is in agreement with what we have found 
in Section \ref{subsec:ism} (the left panel of Figure \ref{fig_o32r23}) that \obfour\ have both 
\oiii/\oii\ ratio and $R23$ index comparable to those of the GPs. 
In contrast to \obfour, the GPs show compact morphology in $g,r,i$ images 
suggesting that they have a compact stellar component \cite{cardamone09}. 
Although there is no detailed study about the extension of the \oiii\ emission line of the GPs yet, 
the compact size of the GPs seen in the $r$-band image, in which the \oiii\ emission line falls, 
may imply the compact size of the \oiii\ emission line as well. 
The unresolved morphology of the GPs in in the $r$-band image suggested that 
the sizes of the \oiii\ emission line should be less than $5$ kpc \citep{cardamone09}. 
Further study of spatial extension of the \oiii\ emission line for the GPs is necessary 
to confirm the above statement. 
The fact that \obfour\ show an extended profile of the \oiii\ emission line over $\sim14$ kpc makes \obfour\ 
different in sizes from the GPs at $z=0.1-0.4$. 
So we call \obfour\ a giant green pea as it shows the high \oiii/\oii\ ratio like the GPs 
and its spatially extended feature of the \oiii\ emission line.

We replot the diagram of the \oiii/\oii\ ratio against the $R23$ index in the right panel of Figure \ref{fig_o32r23} by separating the radial profiles of each \oiii\ blob into 3 components: the central part and two outer regions on both sides of the profile beyond the stellar continuum. 
We use letters ``L", ``C", and ``R" to represent 
the left, central, and right components of each blob, respectively. 
The left component corresponds to the left part of the radial profile shown in Figure \ref{fig_o32profile}. 
The central component is roughly twice the FWHM of the continuum radial profile to ensure that the extended parts are not contaminated by the stellar contribution. 
As seen in the figure, the left and central components of \obtwo\ are close to the plot of the entire object and are located in the region where the star-forming galaxies 
at $z\sim0$ and $z\sim0.7$ are distributed. 
The line ratio of the right sub-region of \obtwo , however, shows higher $R23$ index and 
agrees well with the line ratios of the local SDSS AGNs.  
Originally, \obthree\ shows the high $R23$ index with the large uncertainty. 
When we divide \obthree\ into three components, the extended parts of \obthree\ show slightly lower $R23$ indices, but are still consistent within the large uncertainties. 
The left part of \obthree\ falls closer to the GP region with the \oiii/\oii\ ratio higher than that of the central part, which is in agreement of Figure \ref{fig_o32profile}. 
Meanwhile, the central and right sub-regions are located close to the integrated line ratio. 

For \obone, the left component moves apart from the rest and falls into the GP region 
of the high \oiii/\oii\ ratios and $R23$ indices. 
It can be argued that the extended component of \obone\ does not have only the comparable \oiii/\oii\ line ratio to the GPs as seen in Figure \ref{fig_o32profile}, but it also have the similar $R23$ index. 
Likewise, the locations of all three components of \obfour\ on the \oiii/\oii-$R23$ diagram 
are consistent with those of the GPs. 
In conclusion, one extended part of \obone\ and the entire \obfour\ show the \oiii/\oii\ ratios and the $R23$ indices consistent with those of the compact GPs at $z=0.1-0.4$.

\subsection{Discovery of A Giant Green Pea at $z=0.838$}
\label{subsec:discovery}

\begin{figure}
	\centering
		\includegraphics[width=0.4\textwidth]{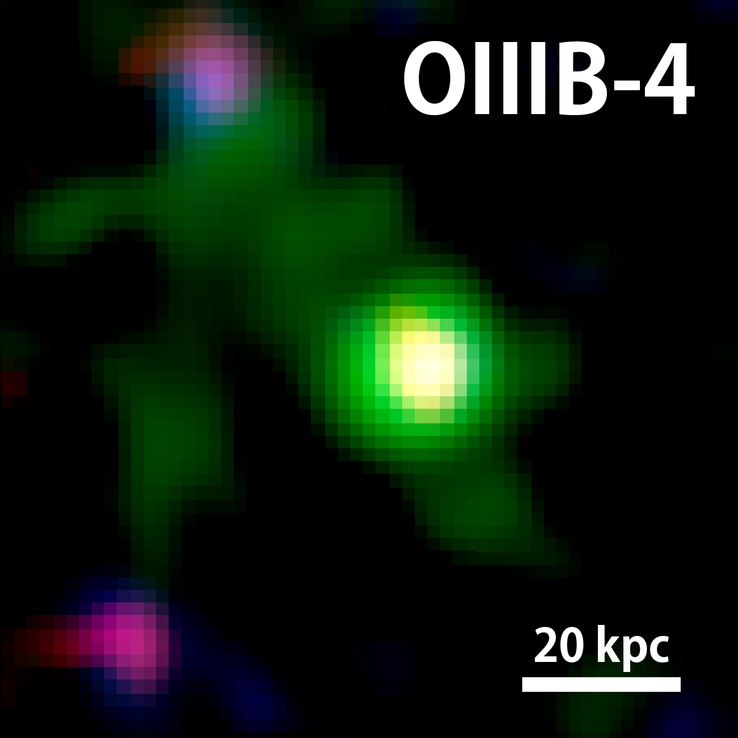}		
	\caption{The composite images of \obfour. North is up; East is left. 
The red and blue colors of both images are the images observed in the 
$R$ and $z$ bands, respectively, 
whereas the green color represents 
the \oiii\ emission line observed in the $NB921$ image. 
The size of each image is roughly $12\times12$ arcsec$^2$, 
which correspond to the physical dimension of $90\times90$ kpc$^2$ at $z=0.8379$. 
The white bar at the bottom right corner indicates the physical scale of $20$ kpc.
}
	\label{fig_rgb}
\end{figure}

We discover a giant GP, \obfour, which has spatially extended \oiii\ emission and 
line ratios consistent with those of the GPs found at $z=0.1-0.4$ by \cite{cardamone09}. 
Figure \ref{fig_rgb} shows the close-up RGB image of \obfour. 
The extended green color in the order of $10-20$ kpc around the center of \obfour\ clearly indicates the large-scale extension of the \oiii\ emission line. 
It is noteworthy that the green components in the North-East (top left) direction of \obfour\ are also parts of \obfour\ when we measure the \oiii\ surface brightness down to $1.2\times10^{-18}$ \ergscmarcsec\ (cf. Figure \ref{fig_spec_target}).

\begin{figure*}
	\centering
	\includegraphics[width=0.45\textwidth]{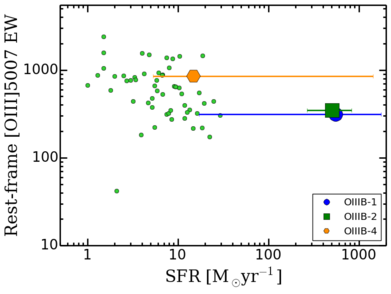}
	\includegraphics[width=0.45\textwidth]{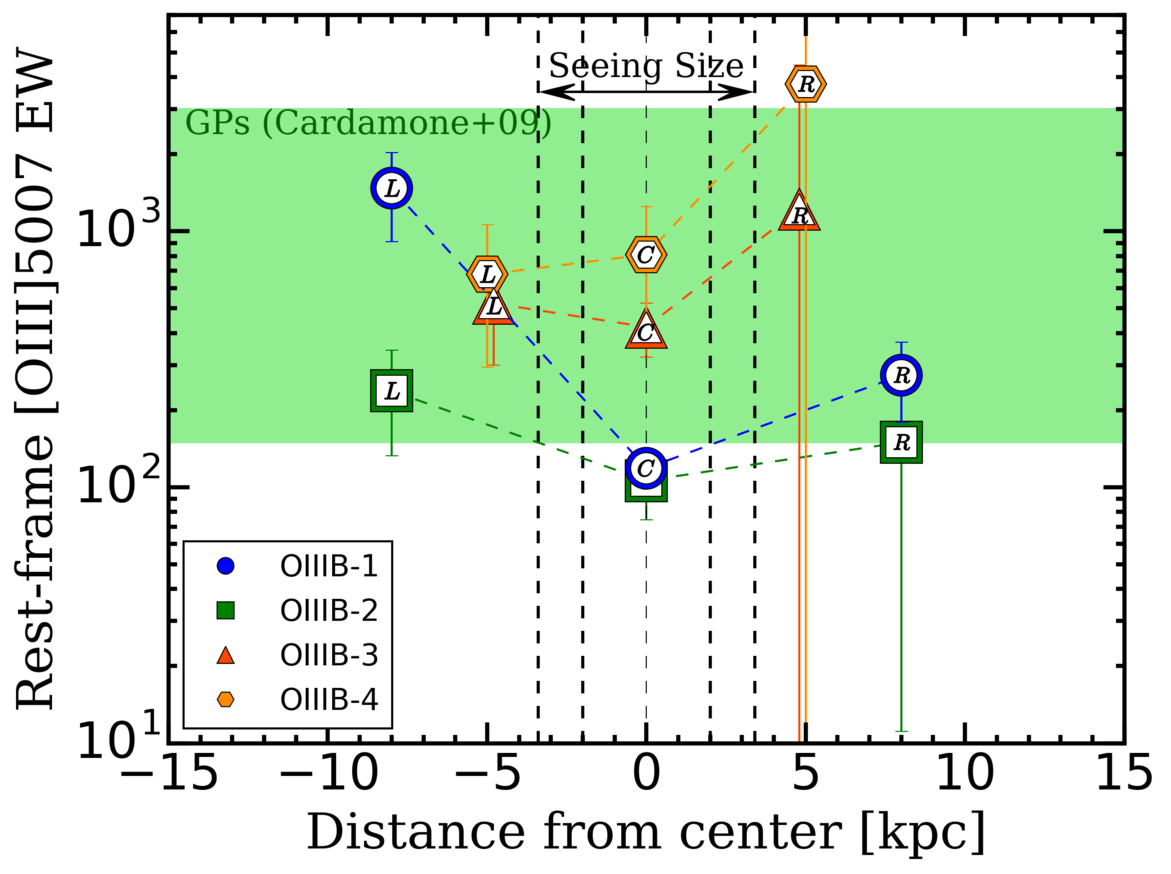}
	\caption{{\it Left panel}: A relation between the rest-frame \oiii\ equivalent widths (\oiii\ EWs) and the star formation rates (SFRs) of 4 \oiii\ blobs at $z\sim0.63$ and $z\sim0.83$. 
	The symbols of the \oiii\ blobs are identical to those in the previous figures. 
	The small green circles show the green peas discovered at $z=0.1-0.4$ by \cite{cardamone09}. 
	{\it Right panel}: The rest-frame \oiii\ EWs of the \oiii\ blobs as a function of the distance from the center. 
	The radial profile of each \oiii\ blob is divided into 3 components: left (L), center (C), and right (R), 
	corresponding to the left, center and right parts of the profile shown in Figure \ref{fig_o32profile}. 	
	}
	\label{fig_ew}
\end{figure*}

The left panel of Figure \ref{fig_ew} shows the relation between the rest-frame \oiii\ equivalent widths 
of the \oiii\ blobs and the SFRs derived by using the SED fitting method. 
In \cite{yuma17}, the observed SEDs of the \oiii\ blobs are collected from the photometry in 9 bands: $BVRizJHK$, IRAC ch1 (3.6\micron), and IRAC ch2 (4.5 \micron)\footnote{IRAC ch1 and ch2 are the mid-infrared data obtained with the Infrared Array Camera from the {\it Spitzer} UKIDSS Ultra Deep Survey (SpUDS; PI: J. Dunlop)}. 
The magnitudes in the $i$ or $z$ band are excluded because they might be contaminated by the strong \oiii\ emission line at $z=0.63$ or $z=0.83$, respectively. 
The stellar population models are constructed with the \cite{bc03} code 
by assuming the constant star formation history, 
the \cite{salpeter55} initial mass function with the mass cutoffs of 0.1 and 100 \Msun, 
\cite{calzetti00} dust attenuation, and the solar metallicity. 
The effect of varying the parameters used to construct the models 
(e.g., the IMF, star formation histories, and metallicity) is included 
in the uncertainties of the derived stellar properties. 
The rest-frame \oiii\ EWs of all \oiii\ blobs are in the ranges of $200-1000$ \AA. 
The giant green pea (\obfour) shows the largest EW of $845\pm27$ \AA. 
We also plot the GPs at $z=0.1-0.4$ in the figure for comparison purpose. 
As mentioned earlier, the GPs are selected as the galaxies with excessive fluxes in the $r$-band image. 
So they should show very high rest-frame \oiii\ EWs by definition. 
The SFRs of the GPs at $z=0.1-0.4$ were derived from the dust-corrected \ha\ fluxes \citep{cardamone09}. 
From the left panel of Figure \ref{fig_ew}, the rest-frame \oiii\ EWs of the GPs are typically 
in the ranges of $100-2000$ \AA. 
All \oiii\ blobs studied in this paper show the EWs in well agreement with the GPs at $z=0.1-0.4$. 
However, \obone and \obtwo\ have significantly higher SFRs than the GPs, indicating 
more intense activity of star formation. 
Meanwhile, the giant green pea (\obfour) show the lower SFR of roughly $15$ \Msun\,yr$^{-1}$, 
which is consistent with the compact GPs at $z=0.1-0.4$. 

We plot the radial profile of the rest-frame \oiii\ EW in the right panel of Figure \ref{fig_ew} 
by dividing the profile into 3 components based on their distances from the center: left ($L$), center ($C$), and right ($R$). 
\obfour\ shows high \oiii\ EWs in all components. The left and central parts of \obfour\ have the \oiii\ EW 
in the ranges of $800-1000$ \AA, consistent with the EW of the entire blob. 
They are in the same ranges as the typical GPs at $z=0.1-0.4$. 
The right component even shows the higher EW, albeit the large error bar due to the low S/N ratio of the faint continuum. 
This clearly confirms that \obfour\ have the \oiii\ EW high enough to be call a giant green pea. 
Among the remaining 3 \oiii\ blobs, \obthree\ shows the high \oiii\ EW comparable to the GPs throughout the entire object. 
On the other hand, \obone\ and \obtwo\ show the \oiii\ EWs at the center slightly smaller than those of the typical GPs at $z=0.1-0.4$. 
Their \oiii\ EWs are in the order of 100 \AA, while the \oiii\ EWs of the extended components increase to approximately 
$150-250$ \AA\ in the case of \obtwo. 
The left component of \obone\ shows the significantly large EW of more than $1000$ \AA. 
It is consistent with what we have found in the right panel of Figure \ref{fig_o32r23} that 
the left part of \obone\ have the \oiii/\oii\ line ratio and the $R23$ index comparable to those of the GPs.

Besides the GPs at $z=0.1-0.4$, there is another population that is similar to our giant green pea. 
\cite{schirmer13} discovered Seyfert-2 galaxies at $z\sim0.3$ in SDSS DR8 with luminous narrow-line regions 
and named these galaxies green beans (GBs). 
GBs show the high \oiii\ luminosities in the order of $10^{43}$ \ergs\ and large \oiii/\oii\ ratios of more than $1.0$. 
Compared with GPs at similar redshifts, GBs are much larger with the size of the \oiii\ emission line of $15-20$ kpc 
and show higher \oiii/\oii\ ratios. 
Their \oiii\ EWs are in the order of $1000$ \AA. 
GBs at $z\sim0.3$ are similar to \obfour\ in terms of the large \oiii\ EW and the spatial extension of the \oiii\ emission line. 
However, the \oiii/\oii\ ratios of GBs are slightly higher than those of \obfour. 
The main difference between \obfour\ and GBs is the energy source; 
GBs are Seyfert type-2 AGNs but \obfour\ is a star-forming galaxy. 
We plot the most prominent GB, J2240-0927, in the right panel of Figure \ref{fig_agn}. 
Its location on the blue diagram is consistent with the distribution of local SDSS AGNs, whereas 
\obfour\ is located close to the local star-forming galaxies on the same figure. 
Additionally, the high-ionization lines such as \nev\ emission are not detected in \obfour. 
The $3\sigma$ flux limit for non-detection is roughly $1.0\times10^{-18}$ \ergscm, which 
provides the $3\sigma$ upper limit on the \nev/\hb\ ratio of \obfour\ to be $<0.03$. 
In contrast, the \nev\ emission line is detected in many of GBs including J2240-0927 
with the \nev/\hb\ ratio of 0.22 \citep{schirmer16}.

\begin{figure*}
	\centering
	\begin{tabular}{cc}
		\includegraphics[width=0.4\textwidth]{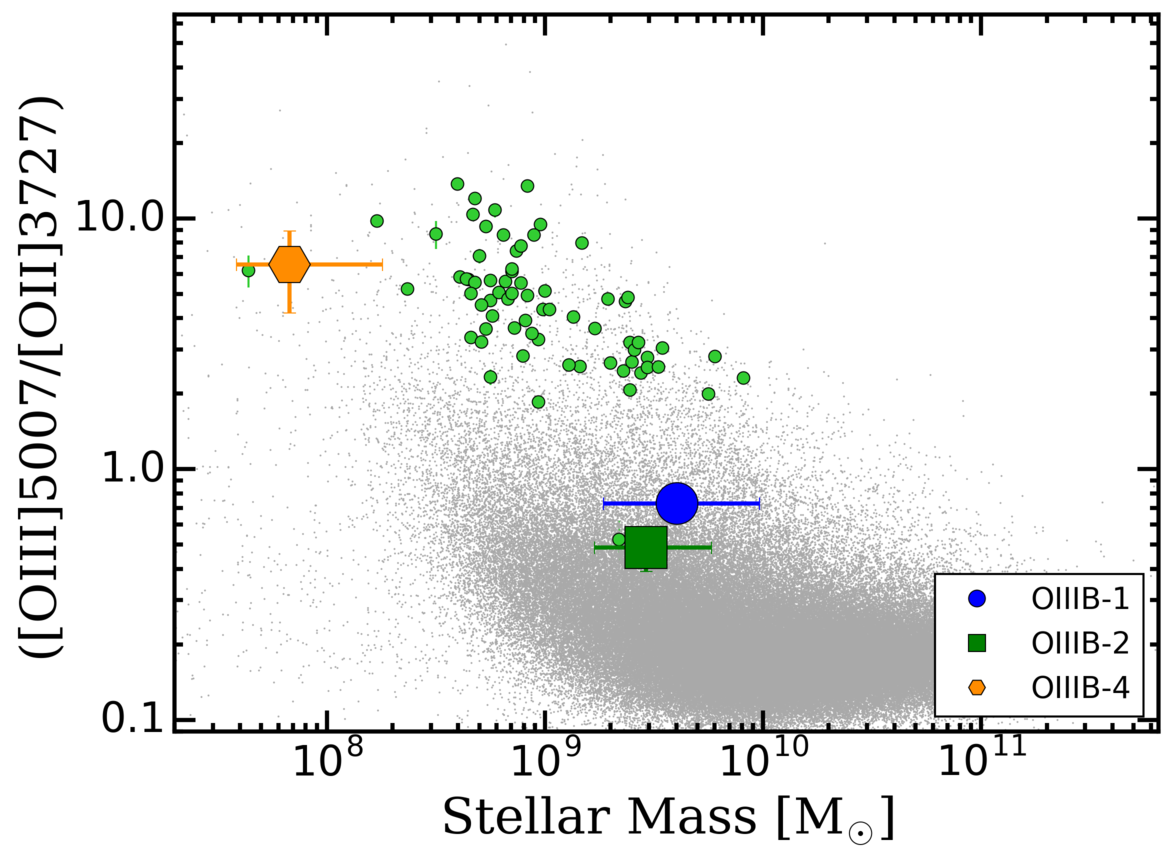} & 
		\includegraphics[width=0.4\textwidth]{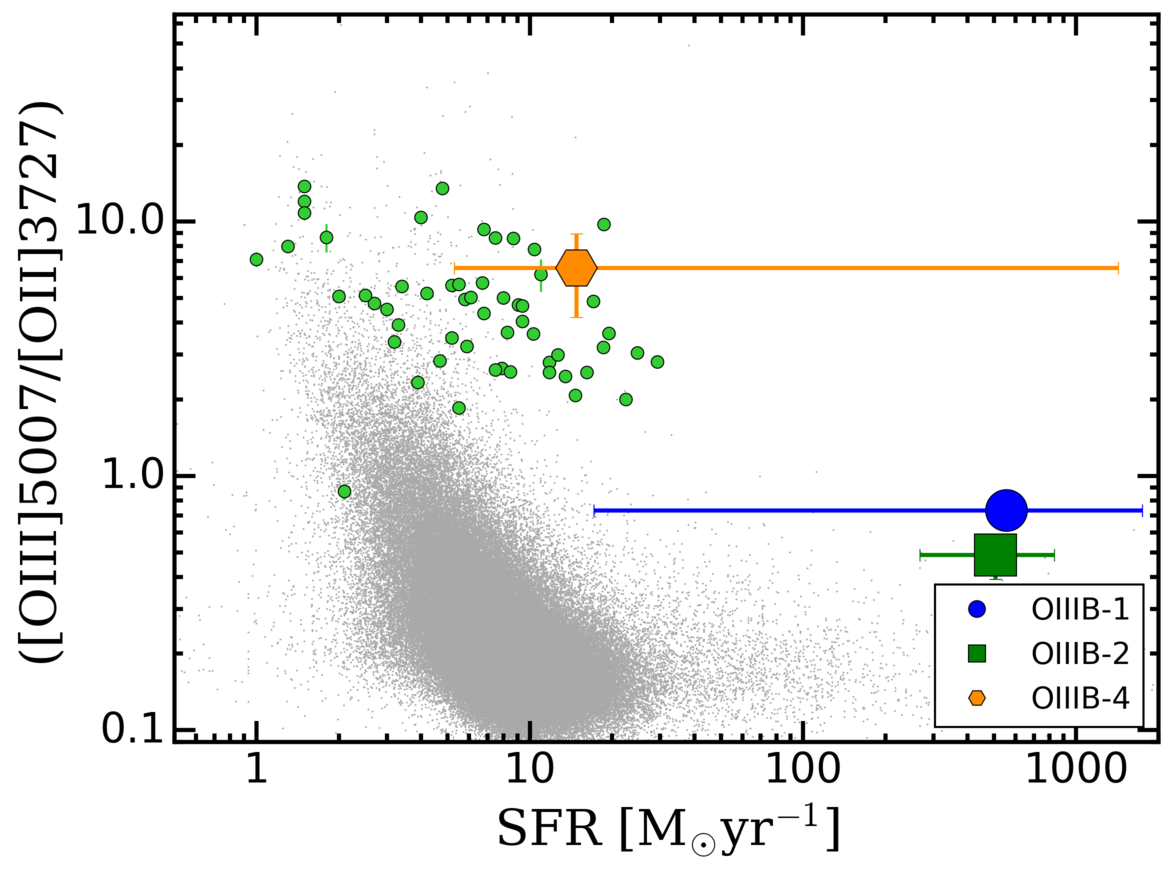} \\
		\includegraphics[width=0.4\textwidth]{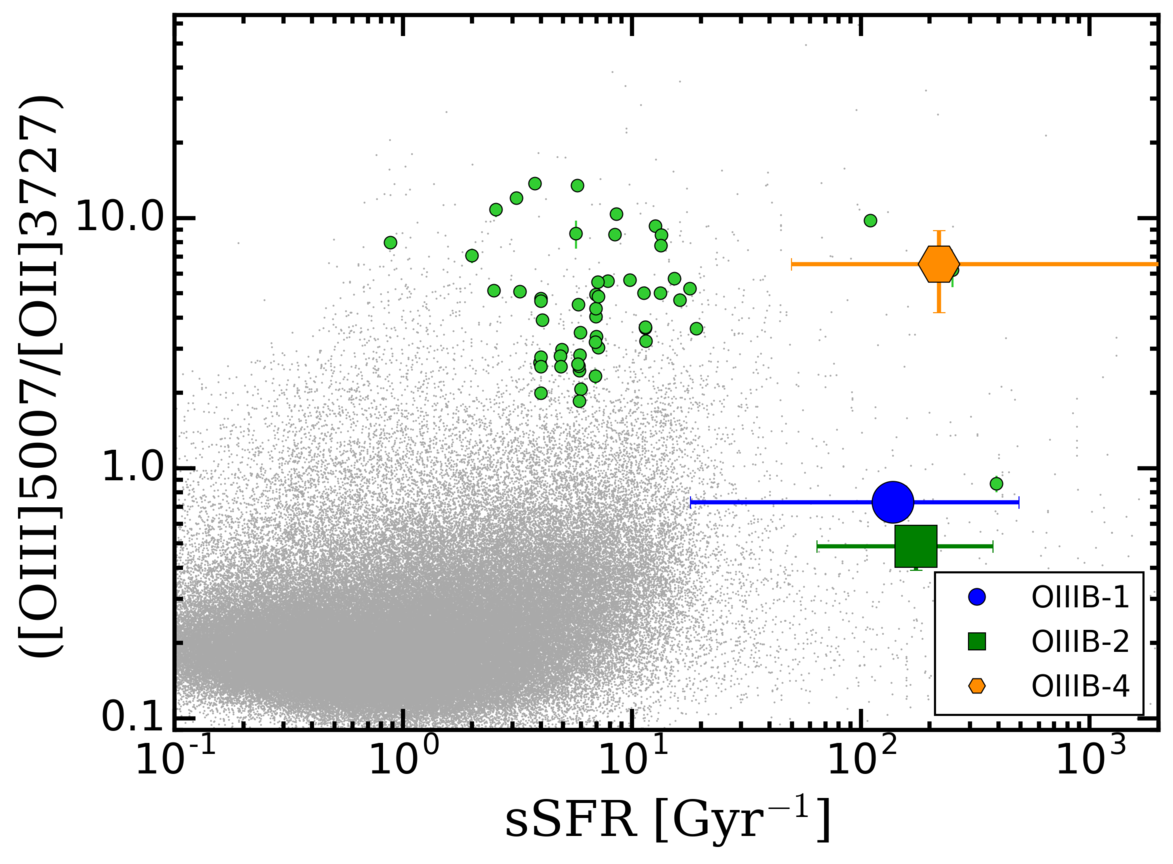} & 
		\includegraphics[width=0.4\textwidth]{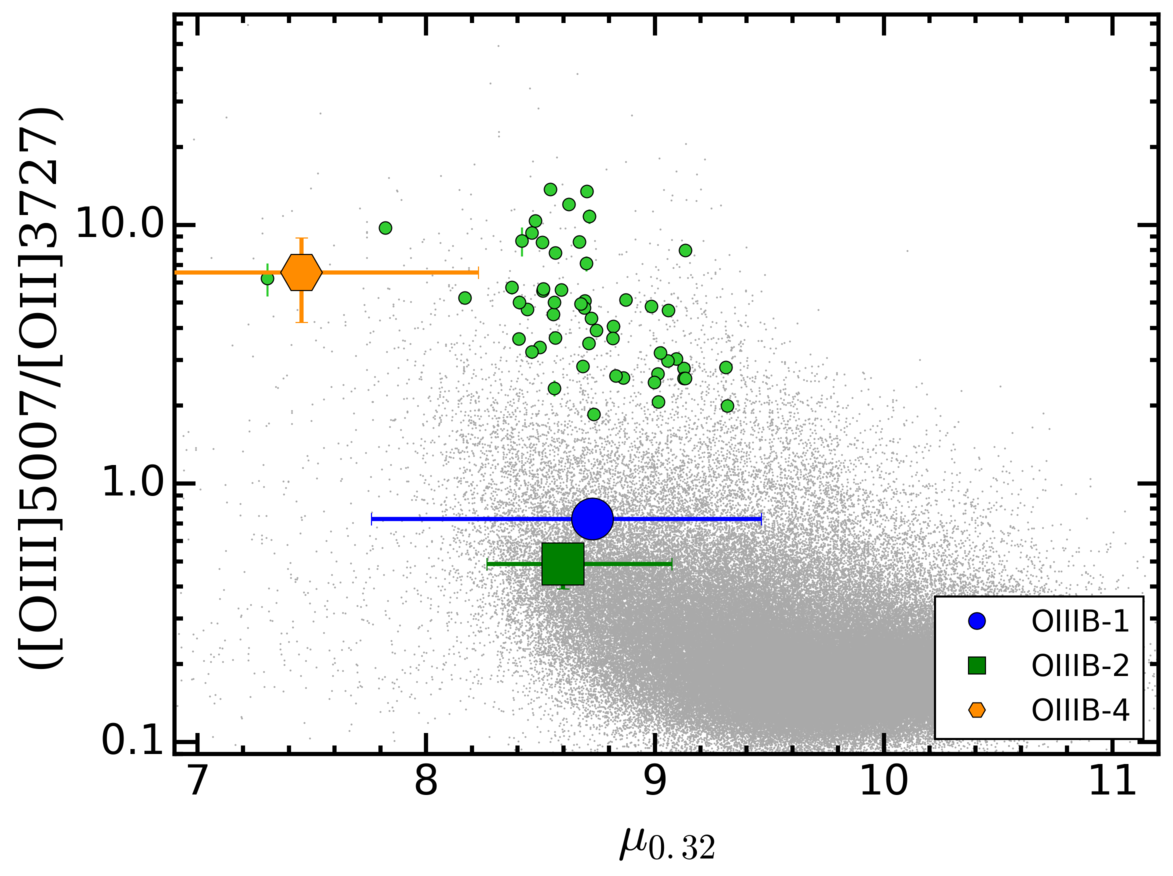}
	\end{tabular}
	\caption{Diagrams of the \oiii/\oii\ ratios versus the stellar mass ($M_{*}$; {\it top left}), 
	star formation rate (SFR; {\it top right}), specific star formation rate (sSFR; {\it bottom left}), 
	and $\mu_{0.32}$ (defined by \cite{mannucci10}; {\it bottom right}). 
	Symbols are identical to those in the previous figures. 
	}\label{fig_o32other}
\end{figure*}

Figure \ref{fig_o32other} shows the correlation between the \oiii/\oii\ line ratios and four different physical properties: the stellar mass ($M_{*}$), star formation rate (SFR), 
specific SFR (sSFR=SFR/$M_{*}$), and $\mu_{0.32}$\footnote{Note that we do not plot \obthree\ in Figure \ref{fig_o32other} because the AGN contribution might affect the SED fitting results.}.  
The stellar mass and SFR of the \oiii\ blobs are derived by the SED fitting method \citep{yuma17}. 
$\mu_{0.32}$ is defined as a combination of $M_{*}$ and the SFR proposed by \cite{mannucci10}: 
\begin{equation}
	\mu_\alpha = \log_{10} (M_*) - \alpha \log_{10} ({\rm SFR}), 
\end{equation}
where $\alpha=0.32$ provides the minimum scatter of the SDSS galaxies in the local universe
in the $\mu_{\alpha}$-metallicity plane. 
The metallicity in terms of $12+\log ({\rm O/H})$ can be empirically estimated from $\mu_{0.32}$ as 
\begin{equation}
	12+\log{\rm (O/H)} = 8.90 + 0.39 x - 0.20x^2 -0.077x^3 +0.064x^4,
	\label{eq:metal}
\end{equation}
where $x=\mu_{0.32}-10$ \citep{mannucci10}. 
It is noteworthy that Equation \ref{eq:metal} is obtained by fitting the local SDSS galaxies with stellar masses of $\log(M_*)=9-11$ \Msun\ or $\mu_{0.32}$ in the range of $\mu_{0.32}=9.0-11.5$. 
In this fitting range, the metallicity $12+\log ({\rm O/H})$ decreases with $\mu_{0.32}$. 
We would reach the minimum metallicity of $12+\log ({\rm O/H})=8.4$ at $\mu_{0.32}=8.73$. 
This relation might not be highly accurate for galaxies with $\mu_{0.32}$ smaller than $8.73$. 
However, it seems that the metallicity continues to decrease at $\mu_{0.32}$ below $8.73$, 
but becomes constant at $\mu_{0.32}$ greater than $10.5$. 
\cite{mannucci10} also proposed that the metallicity of star-forming galaxies with any stellar mass, 
SFR, and at any redshift up to $z=2.5$ practically follows the linear correlation between $12+\log ({\rm O/H})$ 
and $\mu_{0.32}$: 
\begin{equation}
	12+\log{\rm (O/H)} = 8.90 + 0.47 x \hspace{1cm}{\rm if}~\mu_{0.32}<10.2, 
	\label{eq:metal2}
\end{equation}
where $x=\mu_{0.32}-10$. 
Therefore, $\mu_{0.32}$ can basically represent the metallicity of the galaxies with the stellar mass in the range of $\log(M_*)=9-11$ \Msun.

As mentioned earlier, the giant GP, \obfour, 
shows the highest \oiii/\oii\ ratio among all four \oiii\ blobs discussed in this paper. 
It is located in different locations from the rest of \oiii\ blobs in all plots in Figure \ref{fig_o32other}. 
The stellar mass of \obfour\ is $6.75_{-2.91}^{+11.22}\times10^7$ \Msun. 
By compared with the other \oiii\ blobs in our sample, 
\obfour\ is the \oiii\ blobs with the smallest $M_{*}$ and the lowest $\mu_{0.32}$. 
The sSFR of \obfour\ is $\sim2\times 10^2$ Gyr$^{-1}$, which is comparable to the other \oiii\ blobs. 
In fact, most of the \oiii\ blobs found at $z=0.63$ and $z=0.83$ show the high sSFR 
that is greater than the sSFR of typical star-forming galaxies 
on the so-called main sequence of the stellar mass-SFR diagram \citep{elbaz07, yuma17}.  
\obfour\ is the \oiii\ blob at $z\sim0.83$ with the smallest stellar mass. 
Even if we compare \obfour\ with the normal \oiii\ emitters at the same redshift, we would still find that \obfour\ is among the emitters with the smallest stellar mass and highest sSFR \citep[cf. Figures 12 and 13 in][]{yuma17}. 
The high \oiii/\oii\ line ratio of \obfour\ is comparable to the $z=0.1-0.4$ GPs; however, 
their stellar properties are not exactly consistent. 
\obfour\ has the stellar mass smaller than a majority of the GPs and SDSS galaxies. 
The sSFR of \obfour\ is consistent with two GPs at $z=0.1-0.4$, 
while it is higher than those of the remaining GPs. 
The low $\mu_{0.32}$ of \obfour\ as seen in the bottom right panel of Figure \ref{fig_o32other} 
may imply its low metallicity. 
\cite{amorin10} showed that the GPs at $z=0.1-0.4$ have 
the metallicities in the ranges of $7.6<12+\log ({\rm O/H})<8.4$. 
The comparable line ratios between \obfour\ and the GPs may imply that 
the metallicity of \obfour\ is likely similar to that of the GPs. 
The implied sub-solar metallicity of \obfour\ might affect the stellar properties derived by the SED fitting 
method in that the models with sub-solar metallicity give slightly lower SFRs (0.12 dex) than those with 
solar metallicity \citep{yabe09}. However, the effect of varying metallicity of models is already included in 
the uncertainties of the derived properties.

With the small stellar mass and low metallicity, \obfour\ is somehow similar to the galaxies 
with extreme emission lines at low redshifts \citep[e.g.,][]{amorin14, izotov17, yang17}. 
\cite{amorin14} found the galaxies at $0.2 \leq z \leq 0.9$ with high rest-frame \oiii\ EWs of $100-1700$ \AA. 
They also suggested the high ionization parameters of $q\geq 10^8$ \cms, the sSFR of $1-100$ Gyr$^{-1}$, 
and the metallicity of $7.5<12+\log ({\rm O/H})<8.3$, which are similar to \obfour. 
\cite{amorin14} also determined the $\mu_{0.32}$ of their sample to be in the range of $\mu_{0.32}=7.0-8.5$. 
This is also consistent with the $\mu_{0.32}$ of \obfour. 
\cite{izotov17} and \cite{yang17} conducted separate surveys and found the galaxies with extreme \oiii/\oii\ 
emission-line ratios at $z<0.07$ and $z<0.05$, respectively. 
\cite{yang17} proposed the photometric method to select these galaxies and called them the blueberry galaxies, 
which are basically GPs with low masses. 
The blueberry galaxies show low stellar masses and high \oiii/\oii\ ratios comparable to those of \obfour. 
However, the only obvious difference between \obfour\ and these extreme emission-line galaxies 
is the physical size of the emission line. 
These galaxies show the size so compact that some of them are unresolved in the SDSS images \citep{yang17}.

The average properties of \obone\ are indistinguishable from those of \obtwo; 
it is located close to \obtwo\ in all plots in Figure \ref{fig_o32other}. 
Although the \oii/\oii\ line ratios of \obone and \obtwo\ are not as high as those of \obfour\ and GPs, 
they are still somewhat higher than the majority of the local star-forming galaxies. 
According to the figure, \obone and \obtwo\ show 
the moderate masses of roughly $3-5\times10^9$ \Msun, the significantly high SFR in the order of few hundreds \Msun\,yr$^{-1}$, and subsequently the high sSFR as compared with the local star-forming galaxies. 
Their $\mu_{0.32}$ values are less than most star-forming galaxies in the local universe, implying 
the metallicities lower than those of the local star-forming galaxies.

\section{Discussion}\label{sec:discuss}
 
We know from Section \ref{subsec:agn} that an AGN activity is possibly responsible for the spatial extension of the \oiii\ emission line in only one blob, \obthree, but we cannot make a constraint on AGN contribution of \obtwo. 
\obtwo\ can be a star-forming galaxy, composite, or AGN. 
So we do not further discuss about the origins of \obtwo\ and \obthree\ because of the possible AGN contribution. 
The giant GP (\obfour) with the remarkably high \oiii/\oii\ line ratios and \obone\ 
are likely to be heated by the star formation activity with the high sSFR. 
\cite{jaskot13} studied six most extreme GPs at $z=0.1-0.4$ with the highest \oiii/\oii\ ratios. 
They examined possible origins of the \heii$\lambda4686$ emission detected in most of their sample 
including Wolf-Rayet stars, AGNs, and fast radiative shocks. 
Because of the size difference of the \oiii\ emission lines between the GPs at $z=0.1-0.4$ and 
\oiii\ blobs at $z\sim0.7$, 
the physical structures in the extended regions of the \oiii\ blobs may be different from the GPs. 
In this section, we discuss plausible scenarios and physical conditions that maybe able to explain 
the extended \oiii\ emission in \obone\ and \obfour\ 
including an AGN-light echo \citep[e.g.][]{schawinski15, ichikawa18}, fast radiative shocks 
\citep[e.g.][]{kewley01, allen08, shirazi12}, and 
the density bounded system \citep[e.g.][]{nakajima13, nakajima14}. 

\subsection{AGN-Light Echo}

An AGN plays an important role in regulating the galaxy evolution. 
The accretion to the supermassive black hole at the center of the AGN produces 
the significant amount of radiative energy 
that is able to photoionize the interstellar medium of the host galaxy 
and cause large-scale gas outflows from the host galaxy \citep[e.g.,][]{silk98}. 
In this mode of the AGN activity, AGN photoionization overcomes the photoionization from 
the \hii\ region of the host galaxy, 
so the host galaxy can be identified as an AGN \citep[e.g.,][]{baldwin81, veilleux87, lamareille10}. 
It has been long known that an AGN can show luminosity variability \citep{ulrich97}. 
It is possible that the AGN photoionization occurs after the accretion stops because 
the light echo from the past AGN activity takes times to travel across the ISM of the host galaxy \citep[e.g.,][]{schawinski15}. 
In this period of the AGN-light echo, the AGN signature such as the X-ray emission at the center of the galaxy 
is not visible, but we can still observe the extended emission of the galaxy. 
Although \obfour\ and \obone\ do not show any signature of the current AGN activity, 
it is still possible that their extended emission lines are the effect of the past AGN activity. 
If this is the case, \obfour\ and \obone\ should show the AGN-like line ratios in the extended components.

\begin{figure}
	\centering
	\includegraphics[width=0.45\textwidth]{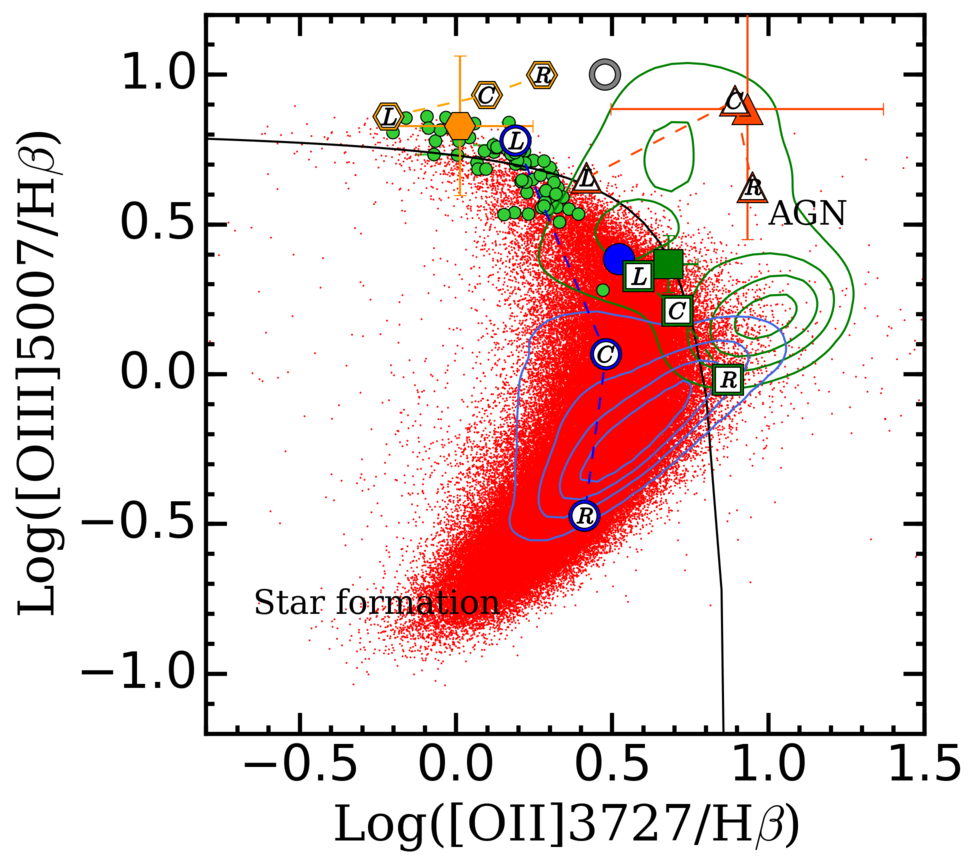}
	\caption{The blue diagram identical to the right panel of Figure \ref{fig_agn} overplotted with three spatial components of the \oiii\ blobs. The letters ``L", ``C", and ``R" correspond to the left, the center, and the right components of each blob. 
The circle, square, hexagon, and triangle represent \obone, \obtwo, \obthree, and \obfour, respectively. 
Small green circles are the green peas at $z=0.112-0.360$ by \cite{cardamone09}.
Red dots are the local SDSS star-forming galaxies. 
Blue and green contours represent composites and AGN, respectively. }
	\label{fig_blue_profile}
\end{figure}

%

We plot the blue diagram again in Figure \ref{fig_blue_profile}, 
but this time we separate the line ratios into 3 components based on the distance from the center of the \oiii\ blob 
(see also Section \ref{subsec:profile}). 
For comparison purpose, we also plot the line ratios of the GPs at $z=0.1-0.4$. 
All components of \obfour\ show the \oiii/\hb\ ratios consistent with the GPs and some fraction of 
the local star-forming galaxies that show the high \oiii/\hb\ ratios. 
For \obone, the left component with the high \oiii/\oii\ ratio seen in the right panel of Figure \ref{fig_o32r23} and in Figure \ref{fig_o32profile} is located in the GP region. 
Therefore, it is unlikely that the extended \oiii\ emission lines of \obfour\ and \obone\ are caused by the AGN 
photoionization left as the AGN-light echo after the super massive black hole at the center stops accreting. 
It is also seen from the figure that all components of \obtwo\ lie in the intersect region of 
star-forming galaxies, composites, and AGNs. The line ratios of the central and right components of \obthree\ are 
consistent with those of AGNs, while the left component shows the lower \oii/\hb\ ratio, 
closer to the star-forming galaxies and the GPs. 

\subsection{Fast Radiative Shock}

Without the AGN contribution, the extended \oiii\ emission line and the gas outflow observed in \obfour\ and \obone\ are most likely fueled by the intense star formation activity (see also Section \ref{subsec:discovery}). 
However, some physical conditions are still required to explain such high \oiii/\oii\ ratios in the extended parts of the \oiii\ blobs. 
Supersonic motion of supernovae or gas outflows from a galaxy is able to create the radiative shock 
with the energy high enough to strongly affect the \hii\ region and the ISM of the galaxy. 
A fast radiative shock, where extreme ultraviolet and soft X-ray photons 
are created by strong ionizing radiation behind the shock front, 
is one of the plausible explanations for the existence of the highly ionized 
emission lines \citep[e.g.,][]{dopita96, allen08}. 

\begin{figure*}
	\centering
	\includegraphics[width=0.8\textwidth]{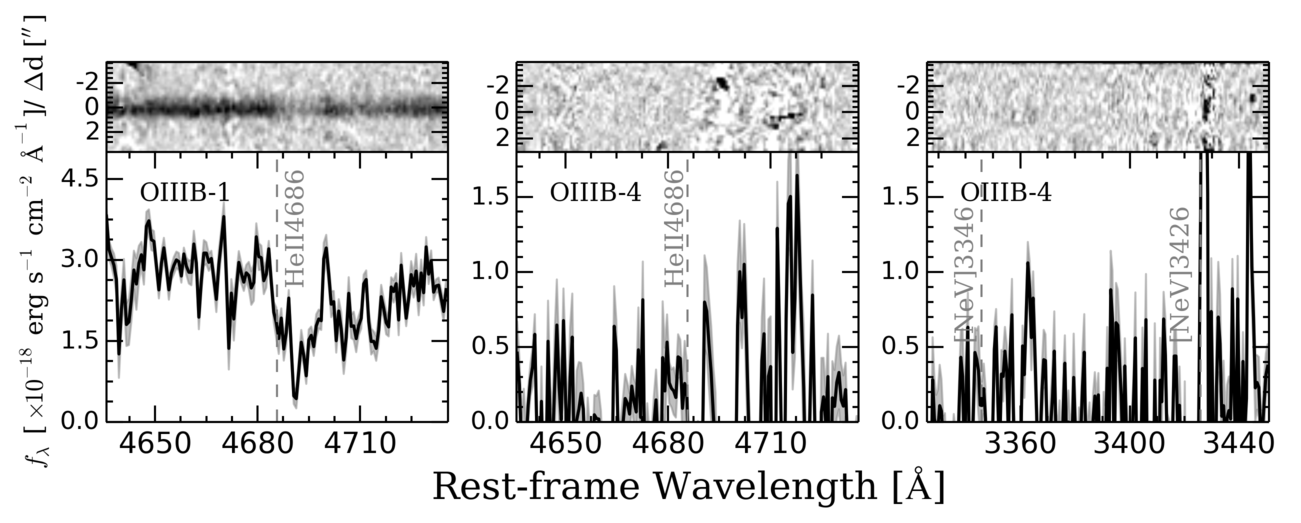}
	\caption{1D and 2D spectra of \obone\ ({\it left panel}) and \obfour\ ({\it middle and right panels}) 
	at the wavelengths of \heii$\lambda4686$ and \nev$\lambda\lambda3346,3426$. 
	The expected wavelength of each line is indicated with the gray dashed line. 
	}\label{fig_heii}
\end{figure*}

\cite{allen08} developed the library of radiative shock models called ``MAPPINGS III," 
for the ISM with various ranges of metallicities (solar, Small and Large Magellanic Cloud metallicities), 
pre-shock densities of $0.01-1000$ cm$^{-3}$, and shock velocities of $100-1000$ \kms. 
When the velocity of the shock increases, the velocity of photoionization front increases 
and exceeds the shock velocities at a certain velocity limit ($v_{\rm shock}\approx 170$ \kms). 
At this point, the photoionization front is separated from the shock front and forms a ``precursor" 
of the \hii\ region, which dominates the optical emission of shocks at high shock velocities \citep{allen08}. 
In the models of shock+precursor, the \oiii/\hb\ ratios become consistent with those of the extended 
components of the \oiii\ blobs; i.e., $\log({\rm O}\,\textsc{III}]/{\rm H}\beta)\geq 0.5$ (see Figure \ref{fig_blue_profile}), 
when the shock velocity reaches $v_{\rm shock}\geq 350$ \kms\ regardless of the values of 
the magnetic field and the metallicity. 
The prominent emission lines that is indicative of the fast radiative shocks include 
\heii$\lambda4686$ and \nev$\lambda\lambda3346,3426$ \citep[e.g.,][]{dopita96, thuan05, izotov12, jaskot13}. 
However, we do not detect the \heii\ or \nev\ emission line in \obfour\ and \obone\ 
above $3\sigma$ flux limit of roughly $1.0\times10^{-18}$ \ergscm. 
This places the upper limit of the \heii/\hb\ ratios to be $<0.03$ in the case of \obfour. 
\obone\ even seems to show the \heii$\lambda4686$ as an absorption line (the left panel of Figure \ref{fig_heii}). 
The non-detection of \heii\ and \nev\ of \obfour\ is also shown in the middle and right panels of the figure, respectively. 
The lowest shock velocity of 100 \kms\ in the MAPPINGS III shock models by \cite{allen08} suggests 
the \heii/\hb\ line ratio of $0.47$ assuming the density of $n=1.0$ cm$^{-3}$ and solar abundance.

\begin{figure*}
	\centering
	\includegraphics[width=0.8\textwidth]{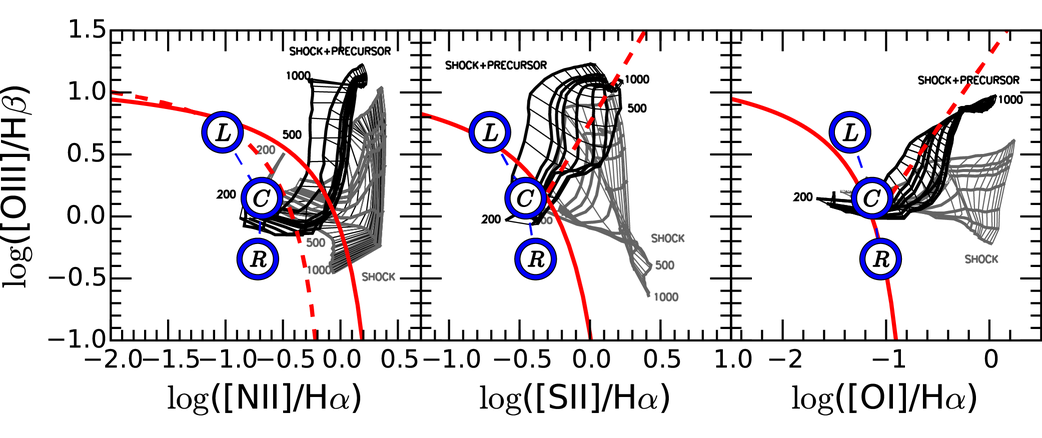}
	\caption{Emission-line diagnostic diagrams of \obone. 
	Symbols for \obone\ are identical to those in the previous figures. 
	The gray and black grids show the shock-only and shock+precursor models 
	with solar abundance and preshock density of $n=1\,{\rm cm}^{-3}$
	obtained from MAPPINGS III \citep{allen08}, respectively. 
	The grid consists of the lines of constant shock velocity in the range of $200-1000$ \kms\ with 50 \kms\ intervals (thin lines) 
	and the lines of constant magnetic parameter ranging from $B/n^{3/2}=0.0001$ $\mu{\rm G}\,{\rm cm}^{3/2}$ 
	to $B/n^{3/2}=10.0$ $\mu{\rm G}\,{\rm cm}^{3/2}$ (thick lines). 
	}
	\label{fig_shock}
\end{figure*}

For \obone\ that we have Keck/MOSFIRE NIR spectrum, we also use another line diagnostics 
to investigate the shock scenario. Figure \ref{fig_shock} shows the line ratios of each component of \obone\ 
in various emission-line diagnostic diagram. 
The grids of both shock-only and shock+precursor models are also illustrated in the figure 
in the case of solar metallicity and preshock density of $n=1\,{\rm cm}^{-3}$. 
It is clearly seen that the left component of \obone\ is not in the shock regions in all diagrams. 
Therefore, it is unlikely that the fast radiative shock is a physical process responsible for 
the high ionization parameters observed in \obfour\ and the extended part of \obone.

\subsection{Density Bounded Nebulae}

As already seen in Figure \ref{fig_o32r23}, 
the \oiii/\oii\ ratio strongly depends on the ionization parameter and metallicity.  
The \oiii/\oii\ ratio increases with increasing ionization parameters and decreasing metallicities. 
In addition, \cite{shirazi14} argued that the high gas density of high-redshift galaxies affects 
the extreme value of \oiii/\oii\ ratios. 
\cite{nakajima14} studied the ISM of star-forming galaxies at $z=0-3$ and 
found that the \oiii/\oii\ line ratio also depends on the escape fraction of the ionizing photons, 
which is defined as a ratio of ionizing photons escaping the galaxy to the totally produced ionizing photons. 
To explain the relationship between the \oiii/\oii\ ratio and the escape fraction, 
they classified star-forming nebulae into 2 types: ionization bounded and density bounded nebulae. 
The size of the ionization-bounded nebulae is equivalent to the Str\"omgren radius, which is determined by the ionization equilibrium between the producing rate of ionizing photons and the recombination rate. 
In the ionization bounded nebulae, high-ionization state of oxygen ions (producing \oiii) locates close to the ionizing sources, while the envelope of the O$^+$ ions that produce \oii\ is around the edge of the nebulae \citep[e.g,][]{shields90, oey97, pellegrini12, nakajima14}. 
On the other hand, some fraction of the ionizing photons is allowed to escape from the homogeneous density bounded nebulae. 
The \hi\ clouds around the density bounded nebulae are so small that they are completely ionized. 
The radius of the density-bounded nebulae is thus constraint by the density of the gas cloud instead of the Str\"omgren radius, as the complete Str\"omgen sphere cannot be formed. 
The size of O$^+$ region is thus smaller in the density bounded nebulae than it is in the ionization bounded system. 
With similar size of the O$^{2+}$ region, the density bounded nebulae would show larger \oiii/\oii\ ratio in the extended part. 

The high \oiii/\oii\ ratios of \obfour\ may be explained by the hypothesis that the \oiii\ blob is the density bounded nebula. The \oiii/\oii\ ratio in the outer region is due to the escaping ionizing photons produced by the intense star formation activity at the center of the blob. 
\cite{arthur11} performed the radiation-magnetohydrodynamic (MHD) simulation of 
the formation and expansion of the \hii\ regions. 
They showed that the highly ionized density-bounded nebula is strongly influenced by 
the radiative feedback and gas instability, leading to the peculiar gas morphologies. 
\obone\ has only one side with high \oiii/\oii\ ratio, implying the asymmetric morphology of the outflow. 
However, it does not explain why the \oiii/\oii\ ratio increases at the greater distance from the center of \obone. 
Normally, the \oiii/\oii\ ratio should decrease as a function of the distance from the center, 
as the O$^{+}$ ions increase while the O$^{2+}$ ions decrease at the outer region \citep{pellegrini12}. 
The studies of diffuse ionized gas (DIG) beyond the \hii\ regions of the local galaxies
typically show the decrease in \oiii/\oii\ ratios at the outer region of the galaxies 
and suggest the lower ionization parameters of DIG than those of the \hii\ regions \citep[e.g.,][]{zhang17}. 
The higher \oiii/\oii\ ratio of the left component of \obone\ than the rest of the galaxy seems 
to contradict with the typical DIG in the local universe. 
Nonetheless, \cite{weber19} performed the numerical analysis to predict the line ratios of \hii\ regions and DIG 
by varying the metallicity, stellar effective temperature, and escape fraction $f_{\rm esc}$. 
They found that the \oiii/\oii\ ratios increase with radius only in the models of 
the density-bounded leaky \hii\ regions with the stellar effective temperature of $40,000$ K 
and the escape fraction less than 10\%. 
This might be the case for \obone. 
\obone\ may have the peculiar gas morphology that can cause the leakage of highly ionized photons through 
a hole in some direction away from the line of sight to the center of the galaxy. 
Dense gas clouds in the line of sight lead to the low escape fraction, whereas the \oiii\ emission 
is largely scattered toward us from the extended region \citep{bassett19}. 
Further constraint on the escape fraction of \obone\ is necessary to confirm this scenario. 

\section{Summary}\label{sec:summary}

\oiii\ blobs are galaxies that exhibit the strong and spatially extended \oiii\ emission line 
beyond their stellar continuum. 
The first systematic survey of \oiii\ blobs has been conducted by \cite{yuma17}. 
In this paper, we investigate the physical nature of the \oiii\ blobs in more details 
by using optical and near-infrared spectroscopy. 
We choose the spectroscopic targets based on 
their large extension of the \oiii\ emission line 
and their coordinates to maximize the number of targets per the observed mask. 
As a result, we observed two largest \oiii\ blobs at $z\sim0.63$ (\obone\ and \obtwo) and another two \oiii\ blobs at $z\sim0.83$ (\obthree\ and \obfour) 
in the optical wavelengths with Subaru/FOCAS. 
\obone\ was also observed in the near-infrared wavelength with Keck/MOSFIRE. 
The main results are listed below. 

\begin{enumerate}
\item We confirm the spectroscopic redshifts of all targets to be at the expected redshifts; i.e., \obone\ and \obtwo\ are at $z=0.6210$ and $z=0.6413$, respectively. \obthree\ and \obfour\ are at $z=0.8365$ and $z=0.8379$, respectively. 

\item The original survey of \oiii\ blobs was based on the hypothesis that the spatially extended \oiii\ emission line is caused by the large-scale outflow of gas from the center of galaxies. 
We could confirm the outflow signature in \obone, 
which has the stellar continuum bright enough to be detected with the high S/N ratios. 
The blueshifted \feii$\lambda2326$ absorption, 
\feiis$\lambda2626$ fine-structure emission, and 
\mgii$\lambda2804$ absorption lines are detected with the central velocity offsets of $-160$ \kms, 
$-200$ \kms, and $-270$ \kms, respectively. 
The outflow velocities are slightly higher than those of normal star-forming galaxies at $z=1-2$ by \cite{erb12}. 

\item We examine AGN contribution by using the BPT diagram for \obone\ with available NIR data and 
the \oiii/\hb\ vs. \oii/\hb\ diagram for the other three blobs, of which we have only the optical spectroscopic data. 
Among 4 observed \oiii\ blobs, \obthree\ is the only one that is heated by an AGN activity at the center 
based on its location on the \oiii/\hb\ vs. \oii/\hb\ diagram. 
\obone, on the other hand, is located in the region consistent with the the pure star formation activity on the BPT diagram. 
\obfour\ is another \oiii\ blob that is identified as a normal star-forming galaxies in the blue diagram, though it shows a remarkably high \oiii/\oii\ line ratio. 
Lastly, we could not rule out the possibility of AGN contribution in \obtwo; it is likely to be a composite powered by both star formation and AGN activities. 

\item The properties of the interstellar medium surrounding the \oiii\ blobs have been examined by 
the \oiii/\oii-$R23$ diagram. 
\obone, \obtwo, and \obthree\ show the \oiii/\oii\ line ratios comparable to those of the local SDSS galaxies and the star-forming galaxies at $z\sim0.7$. 
Their inferred ionization parameters should be slightly less than $q_{ion}=8\times10^7$ \cms. 
Interestingly, \obfour\ exhibits the impressively high \oiii/\oii\ ratio. The line ratio and the $R23$ index of \obfour\ agree well with those of the compact green peas at $z=0.1-0.4$. 
The ionization parameter of \obfour\ should be roughly in the order of $10^8$ \cms.

\item The radial profile of the \oiii/\oii\ ratios indicates that \obfour\ shows the \oiii/\oii\ ratio much greater than those of the star-forming galaxies at $z\sim0.7$ and the other 3 \oiii\ blobs. 
The radial profile of \obfour\ shows the almost constant \oiii/\oii\ emission-line ratios of approximately $5-10$ over $14$ kpc. 
The \oiii/\oii\ ratios are consistent with those of the GPs at $z=0.1-0.4$; however, the physical sizes are different as the GPs are compact with 
the \oiii\ emission lines within $5$ kpc. 
Thus we call \obfour\ a giant green pea. 
Among the remaining 3 blobs, there is only one extended part of \obone\ showing the \oiii/\oii\ ratio as high as the GPs. 
It is suggested that such a high \oiii/\oii\ ratio found in the GPs at $z=0.1-0.4$ can also be found in the extended component beyond the stellar continuum of the galaxies as well. 

\item The rest-frame \oiii\ equivalent width of \obfour\ is $845\pm27$ \AA. 
This is well in agreement with the typical GPs at $z=0.1-0.4$, 
which are the galaxies with high \oiii\ EWs by definition. 
The \oiii\ EWs of \obfour\ are still consistent with those of the GPs, 
whose \oiii\ EWs are in the range of $100-2000$ \AA, 
even if we divide the radial profile into 3 components according to the distance from the center of the blob. 
\obone, \obtwo, and \obthree\ also have the rest-frame \oiii\ EWs in the same ranges as the GPs. 
When we divide the EWs into 3 components, the large \oiii\ EW of the left component of \obone\ 
($>1000$ \AA) is obviously seen. 
It is in agreement with the \oiii/\oii\ ratio and the $R23$ index of this components that are similar to those of the GPs. 
However, the centres of \obone\ and \obtwo\ show the \oiii\ EWs of $\sim100$ \AA\ slightly smaller than those of the GPs. 

\item The giant green pea, \obfour, is also similar to green beans at $z\sim0.3$ in that 
they show comparable \oiii\ EWs and consistent spatial extension of the \oiii\ emission line. 
However, unlike \obfour, GBs is powered by AGN rather than star formation.

\item The giant GP, \obfour, is the \oiii\ blob at $z\sim0.83$ that has the low stellar mass of $7\times10^7$ \Msun, the high specific SFR of $2\times10^2$ Gyr$^{-1}$, and the low metallicity. 
Its physical properties are similar to those of the GPs found at $z=0.1-0.4$.

\item We check whether or not the \oiii\ extension of \obfour\ and \obone\ are an AGN-light echo influenced by the AGN activities in the past. 
The blue diagram suggests that the extended parts of \obfour\ and \obone\ are caused by the star-forming activity. 
Thus we can rule out the AGN-light echo scenario. 

\item The fast radiative shock is another potential mechanism to explain the high \oiii/\oii\ ratios. 
We do not detect the shock-indicative emission lines such as \nev$\lambda\lambda3346,3426$ and \heii$\lambda4686$ in \obfour\ and \obone\ at the $3\sigma$ flux limit of $1.0\times10^{-18}$ \ergscm. 
The upper limit of the \heii/\hb\ ratios ($<0.03$) suggests that the fast radiative shock is unlikely. 

\end{enumerate}

The most likely scenario to explain the high \oiii/\oii\ ratios observed in \obfour\ and the extended component of \obone\ is probably density bounded nebulae. In the case of the density bounded nebulae, high ionization photons can escape from the center of the galaxy and the O$^{+}$ region producing \oii\ emission line is small because of the small \hi\ cloud. 
The density bounded region at the center might lead to a harder emitted spectrum that ionizes the more extended nebulae. 
Further observations are needed to confirm if the physical conditions of the \oiii\ blobs are really consistent with the density bounded nebulae. \\\\

The authors are very grateful to an anonymous referee for valuable comments 
that help improve the article. 
This work is supported by Faculty of Science, Mahidol University, Thailand and the Thailand Research Fund (TRF) through a research grant for new scholar (MRG6180279). We thank Moire Prescott for her useful comments. 
S.Y. thanks David J. Ruffolo for supports as a mentor of the TRF research grant. M.O. is supported by World Premier International Research Center Initiative (WPI Initiative), MEXT, Japan, and KAKENHI (15H02064, 17H01110, and 17H01114) Grant-in-Aid for Scientific Research (A) through Japan Society for the Promotion of Science. 

\appendix
\section{Correction for Dust Attenuation}
\label{appen:dust}
In order to obtain accurate results of the observed fluxes and the line ratios, 
we need to take dust extinction into account. 
For \obone\ and \obtwo, we are able to detect the \hg\ emission lines and use the Balmer decrement to derive the dust attenuation. In addition, we can also detect the \ha\ emission line of \obone\ from the Keck/MOSFIRE spectrum. 
The observed \hg, \hb, and \ha\ fluxes of \obone\ are $4.92\times10^{-17}$ \ergscm, $1.24\times10^{-16}$ \ergscm, and $4.76\times10^{-16}$ \ergscm, respectively. 
These result in the \hg/\hb\ and \ha/\hb\ line ratios of $0.40\pm0.08$ and $3.84\pm0.40$, respectively. 
The \hg/\hb\ line ratio for \obtwo\ is $0.39\pm0.26$ because the \hg\ and \hb\ emission lines show the observed fluxes 
of $1.04\times10^{-16}$ \ergscm, and $2.70\times10^{-16}$ \ergscm, respectively. 
The color excess $E(B-V)$ can be derived from the \hg/\hb\ ratio as follow. 

\begin{eqnarray}
	E(B-V) &=& \frac{E(H\beta-H\gamma)}{\kappa(H\beta)-\kappa(H\gamma)}\\
&=& \frac{-2.5}{\kappa(H\beta)-\kappa(H\gamma)}\times \log_{10}\left[\frac{0.469}{(H\gamma/H\beta)_{\rm obs}}\right].
\end{eqnarray}

Likewise, the \ha/\hb\ line ratio is related to the color excess as 
\begin{eqnarray}
	E(B-V) &=& \frac{E(H\beta-H\alpha)}{\kappa(H\beta)-\kappa(H\alpha)}\\
&=& \frac{-2.5}{\kappa(H\beta)-\kappa(H\alpha)}\times \log_{10}\left[\frac{2.86}{(H\alpha/H\beta)_{\rm obs}}\right].
\end{eqnarray}

The intrinsic \hg/\hb\ and \ha/\hb\ ratios are $0.469$ and $2.86$, respectively. 
They are obtained by assuming Case B recombination with 
an electron density of $10^2$ cm$^{-3}$ and temperature of $10^4$ K \citep{osterbrock89}. 
The reddening curve $\kappa(\lambda)$ is derived by adopting the expression in \cite{calzetti00}. 
Substituting $\kappa($\hg$)=5.12$, $\kappa($\hb$)=4.60$, and $\kappa($\ha$)=3.33$ into the above equations, we obtain the color excesses $E(B-V)$ of $0.33\pm0.19$ mag and $0.24\pm0.10$ mag for \obone\ as derived by using the \hg/\hb\ and \ha/\hb\ line ratios, respectively. They are consistent with each other within $1\sigma$ uncertainty. 
For \obone, we adopt $E(B-V)=0.33$ to correct for the dust attenuation. Because the \hg\ and \hb\ lines are observed simultaneously with the same instrument, they are less suffered from different slit correction or flux calibration as compared to the \ha\ emission line. 
The color excess of \obtwo\ estimated from the \hg/\hb\ ratio is $0.41\pm0.42$ mag. 
These nebular color excesses derived from the Balmer decrement are roughly consistent with those of stellar continuum estimated by the SED fitting method \citep{yuma17} after multiplying by a factor of 0.44 described in \cite{calzetti00}. 
For \obthree\ and \obfour\ whose the \hg\ emission lines are not detected, we estimate the nebular color excesses from the stellar color excesses obtained by the SED fitting method \citep{yuma17}. 
The color excesses of all four blobs are listed in Table \ref{tab_ebv}. 
It is noteworthy that we assumed the constant dust extinction across the \oiii\ blobs. 
We checked the \ha/\hb-ratio profile of \obone\ and found that the \ha/\hb\ ratios decrease slightly with increasing distance up to $5-6$ kpc from the center of \obone. However, the ratios are still within the line-ratio uncertainty of the entire blob. 

\begin{deluxetable}{l rrr}[!h]
\tabletypesize{\small}
\tablewidth{0pt}
\tablecolumns{4}
\tablecaption{Color excesses of 4 \oiii\ blobs at $z=0.63-0.83$ dervied from the Balmer decrement and SED fitting method\label{tab_ebv}}
\tablewidth{0pt}
\tablehead{
\multicolumn{1}{c}{\oiii\ blobs} &
\multicolumn{2}{c}{$E(B-V)_{\rm nebular}$} &
\multicolumn{1}{r}{$E(B-V)_{\rm stellar}$} \\[0.1cm]
\multicolumn{1}{c}{} & 
\multicolumn{1}{c}{\hg/\hb} & 
\multicolumn{1}{c}{\ha/\hb} & 
\multicolumn{1}{r}{SED fitting} 
}
\startdata
  \obone 	 & $0.33\pm0.19$	 & $0.24\pm0.10$ & $0.18_{-0.04}^{+0.34}$ \\[0.1cm]
  \obtwo 	 & $0.41\pm0.42$ & $-$ & $0.20_{-0.20}^{+0.40}$ \\[0.1cm]
  \obthree 	& $-$ & $-$ & $0.34_{-0.20}^{+0.36}$ \\[0.1cm]
  \obfour  	& $-$ & $-$ & $0.18_{-0.10}^{+0.28}$
\enddata
\end{deluxetable}

\vspace{0.5cm}
\bibliographystyle{apj}
\bibliography{extendedGP_new}

\begin{thebibliography}{98}
\expandafter\ifx\csname natexlab\endcsname\relax\def\natexlab#1{#1}\fi

\bibitem[{{Abazajian} {et~al.}(2009){Abazajian}, {Adelman-McCarthy},
  {Ag{\"u}eros}, {et~al.}}]{abazajian09}
{Abazajian}, K.~N., {Adelman-McCarthy}, J.~K., {Ag{\"u}eros}, M.~A., {et~al.}
  2009, \apjs, 182, 543

\bibitem[{{Aguirre} {et~al.}(2008){Aguirre}, {Dow-Hygelund}, {Schaye}, \&
  {Theuns}}]{aguirre08}
{Aguirre}, A., {Dow-Hygelund}, C., {Schaye}, J., \& {Theuns}, T. 2008, \apj,
  689, 851

\bibitem[{{Alexander} {et~al.}(2010){Alexander}, {Swinbank}, {Smail},
  {McDermid}, \& {Nesvadba}}]{alexander10}
{Alexander}, D.~M., {Swinbank}, A.~M., {Smail}, I., {McDermid}, R., \&
  {Nesvadba}, N.~P.~H. 2010, \mnras, 402, 2211

\bibitem[{{Allen} {et~al.}(2008){Allen}, {Groves}, {Dopita}, {Sutherland}, \&
  {Kewley}}]{allen08}
{Allen}, M.~G., {Groves}, B.~A., {Dopita}, M.~A., {Sutherland}, R.~S., \&
  {Kewley}, L.~J. 2008, \apjs, 178, 20

\bibitem[{{Amor{\'\i}n} {et~al.}(2014){Amor{\'\i}n}, {Sommariva}, {Castellano},
  {et~al.}}]{amorin14}
{Amor{\'\i}n}, R., {Sommariva}, V., {Castellano}, M., {et~al.} 2014, \aap, 568,
  L8

\bibitem[{{Amor{\'{\i}}n} {et~al.}(2010){Amor{\'{\i}}n}, {P{\'e}rez-Montero},
  \& {V{\'{\i}}lchez}}]{amorin10}
{Amor{\'{\i}}n}, R.~O., {P{\'e}rez-Montero}, E., \& {V{\'{\i}}lchez}, J.~M.
  2010, \apjl, 715, L128

\bibitem[{{Arthur} {et~al.}(2011){Arthur}, {Henney}, {Mellema}, {de Colle}, \&
  {V{\'a}zquez-Semadeni}}]{arthur11}
{Arthur}, S.~J., {Henney}, W.~J., {Mellema}, G., {de Colle}, F., \&
  {V{\'a}zquez-Semadeni}, E. 2011, \mnras, 414, 1747

\bibitem[{{Baldwin} {et~al.}(1981){Baldwin}, {Phillips}, \&
  {Terlevich}}]{baldwin81}
{Baldwin}, J.~A., {Phillips}, M.~M., \& {Terlevich}, R. 1981, \pasp, 93, 5

\bibitem[{{Bassett} {et~al.}(2019){Bassett}, {Ryan-Weber}, {Cooke},
  {et~al.}}]{bassett19}
{Bassett}, R., {Ryan-Weber}, E.~V., {Cooke}, J., {et~al.} 2019, \mnras, 483,
  5223

\bibitem[{{Bell} {et~al.}(2003){Bell}, {McIntosh}, {Katz}, \&
  {Weinberg}}]{bell03}
{Bell}, E.~F., {McIntosh}, D.~H., {Katz}, N., \& {Weinberg}, M.~D. 2003, \apjs,
  149, 289

\bibitem[{{Benson} {et~al.}(2003){Benson}, {Bower}, {Frenk},
  {et~al.}}]{benson03}
{Benson}, A.~J., {Bower}, R.~G., {Frenk}, C.~S., {et~al.} 2003, \apj, 599, 38

\bibitem[{{Bradshaw} {et~al.}(2013){Bradshaw}, {Almaini}, {Hartley},
  {et~al.}}]{bradshaw13}
{Bradshaw}, E.~J., {Almaini}, O., {Hartley}, W.~G., {et~al.} 2013, \mnras

\bibitem[{{Brammer} {et~al.}(2013){Brammer}, {van Dokkum}, {Illingworth},
  {et~al.}}]{brammer13}
{Brammer}, G.~B., {van Dokkum}, P.~G., {Illingworth}, G.~D., {et~al.} 2013,
  \apjl, 765, L2

\bibitem[{{Bruzual} \& {Charlot}(2003)}]{bc03}
{Bruzual}, G. \& {Charlot}, S. 2003, \mnras, 344, 1000

\bibitem[{{Calzetti} {et~al.}(2000){Calzetti}, {Armus}, {Bohlin},
  {et~al.}}]{calzetti00}
{Calzetti}, D., {Armus}, L., {Bohlin}, R.~C., {et~al.} 2000, \apj, 533, 682

\bibitem[{{Cardamone} {et~al.}(2009){Cardamone}, {Schawinski}, {Sarzi},
  {et~al.}}]{cardamone09}
{Cardamone}, C., {Schawinski}, K., {Sarzi}, M., {et~al.} 2009, \mnras, 399,
  1191

\bibitem[{{Cheung} {et~al.}(2016){Cheung}, {Bundy}, {Cappellari},
  {et~al.}}]{cheung16}
{Cheung}, E., {Bundy}, K., {Cappellari}, M., {et~al.} 2016, \nat, 533, 504

\bibitem[{{Cicone} {et~al.}(2014){Cicone}, {Maiolino}, {Sturm},
  {et~al.}}]{cicone14}
{Cicone}, C., {Maiolino}, R., {Sturm}, E., {et~al.} 2014, \aap, 562, A21

\bibitem[{{Coil} {et~al.}(2011){Coil}, {Weiner}, {Holz}, {et~al.}}]{coil11}
{Coil}, A.~L., {Weiner}, B.~J., {Holz}, D.~E., {et~al.} 2011, \apj, 743, 46

\bibitem[{{Dopita} \& {Sutherland}(1996)}]{dopita96}
{Dopita}, M.~A. \& {Sutherland}, R.~S. 1996, \apjs, 102, 161

\bibitem[{{Elbaz} {et~al.}(2007){Elbaz}, {Daddi}, {Le Borgne},
  {et~al.}}]{elbaz07}
{Elbaz}, D., {Daddi}, E., {Le Borgne}, D., {et~al.} 2007, \aap, 468, 33

\bibitem[{{Erb} {et~al.}(2012){Erb}, {Quider}, {Henry}, \& {Martin}}]{erb12}
{Erb}, D.~K., {Quider}, A.~M., {Henry}, A.~L., \& {Martin}, C.~L. 2012, \apj,
  759, 26

\bibitem[{{F{\"o}rster Schreiber} {et~al.}(2009){F{\"o}rster Schreiber},
  {Genzel}, {Bouch{\'e}}, {et~al.}}]{forster09}
{F{\"o}rster Schreiber}, N.~M., {Genzel}, R., {Bouch{\'e}}, N., {et~al.} 2009,
  \apj, 706, 1364

\bibitem[{{F{\"o}rster Schreiber} {et~al.}(2014){F{\"o}rster Schreiber},
  {Genzel}, {Newman}, {et~al.}}]{forster14}
{F{\"o}rster Schreiber}, N.~M., {Genzel}, R., {Newman}, S.~F., {et~al.} 2014,
  \apj, 787, 38

\bibitem[{{Fotopoulou} {et~al.}(2016){Fotopoulou}, {Buchner},
  {Georgantopoulos}, {et~al.}}]{fotopoulou16}
{Fotopoulou}, S., {Buchner}, J., {Georgantopoulos}, I., {et~al.} 2016, \aap,
  587, A142

\bibitem[{{Fumagalli} {et~al.}(2011){Fumagalli}, {O'Meara}, \&
  {Prochaska}}]{fumagalli11}
{Fumagalli}, M., {O'Meara}, J.~M., \& {Prochaska}, J.~X. 2011, Sci, 334, 1245

\bibitem[{{Genzel} {et~al.}(2011){Genzel}, {Newman}, {Jones},
  {et~al.}}]{genzel11}
{Genzel}, R., {Newman}, S., {Jones}, T., {et~al.} 2011, \apj, 733, 101

\bibitem[{{Harikane} {et~al.}(2014){Harikane}, {Ouchi}, {Yuma},
  {et~al.}}]{harikane14}
{Harikane}, Y., {Ouchi}, M., {Yuma}, S., {et~al.} 2014, \apj, 794, 129

\bibitem[{{Heckman} {et~al.}(1990){Heckman}, {Armus}, \& {Miley}}]{heckman90}
{Heckman}, T.~M., {Armus}, L., \& {Miley}, G.~K. 1990, \apjs, 74, 833

\bibitem[{{Heckman} {et~al.}(2000){Heckman}, {Lehnert}, {Strickland}, \&
  {Armus}}]{heckman00}
{Heckman}, T.~M., {Lehnert}, M.~D., {Strickland}, D.~K., \& {Armus}, L. 2000,
  \apjs, 129, 493

\bibitem[{{Ichikawa} {et~al.}(2018){Ichikawa}, {Ueda}, {Bae},
  {et~al.}}]{ichikawa18}
{Ichikawa}, K., {Ueda}, J., {Bae}, H.-J., {et~al.} 2018, ArXiv e-prints

\bibitem[{{Izotov} {et~al.}(2017){Izotov}, {Thuan}, \& {Guseva}}]{izotov17}
{Izotov}, Y.~I., {Thuan}, T.~X., \& {Guseva}, N.~G. 2017, \mnras, 471, 548

\bibitem[{{Izotov} {et~al.}(2012){Izotov}, {Thuan}, \& {Privon}}]{izotov12}
{Izotov}, Y.~I., {Thuan}, T.~X., \& {Privon}, G. 2012, \mnras, 427, 1229

\bibitem[{{Jansen} {et~al.}(2001){Jansen}, {Lumb}, {Altieri},
  {et~al.}}]{jansen01}
{Jansen}, F., {Lumb}, D., {Altieri}, B., {et~al.} 2001, \aap, 365, L1

\bibitem[{{Jaskot} \& {Oey}(2013)}]{jaskot13}
{Jaskot}, A.~E. \& {Oey}, M.~S. 2013, \apj, 766, 91

\bibitem[{{Kashikawa} {et~al.}(2002){Kashikawa}, {Aoki}, {Asai},
  {et~al.}}]{kashikawa02}
{Kashikawa}, N., {Aoki}, K., {Asai}, R., {et~al.} 2002, \pasj, 54, 819

\bibitem[{{Kauffmann} {et~al.}(2003){Kauffmann}, {Heckman}, {Tremonti},
  {et~al.}}]{kauffmann03}
{Kauffmann}, G., {Heckman}, T.~M., {Tremonti}, C., {et~al.} 2003, \mnras, 346,
  1055

\bibitem[{{Kewley} \& {Dopita}(2002)}]{kewley02}
{Kewley}, L.~J. \& {Dopita}, M.~A. 2002, \apjs, 142, 35

\bibitem[{{Kewley} {et~al.}(2001){Kewley}, {Dopita}, {Sutherland}, {Heisler},
  \& {Trevena}}]{kewley01}
{Kewley}, L.~J., {Dopita}, M.~A., {Sutherland}, R.~S., {Heisler}, C.~A., \&
  {Trevena}, J. 2001, \apj, 556, 121

\bibitem[{{Kewley} {et~al.}(2006){Kewley}, {Groves}, {Kauffmann}, \&
  {Heckman}}]{kewley06}
{Kewley}, L.~J., {Groves}, B., {Kauffmann}, G., \& {Heckman}, T. 2006, \mnras,
  372, 961

\bibitem[{{Kornei} {et~al.}(2012){Kornei}, {Shapley}, {Martin},
  {et~al.}}]{kornei12}
{Kornei}, K.~A., {Shapley}, A.~E., {Martin}, C.~L., {et~al.} 2012, \apj, 758,
  135

\bibitem[{{Lamareille}(2010)}]{lamareille10}
{Lamareille}, F. 2010, \aap, 509, A53

\bibitem[{{Lilly} {et~al.}(2013){Lilly}, {Carollo}, {Pipino}, {Renzini}, \&
  {Peng}}]{lilly13}
{Lilly}, S.~J., {Carollo}, C.~M., {Pipino}, A., {Renzini}, A., \& {Peng}, Y.
  2013, \apj, 772, 119

\bibitem[{{Lilly} {et~al.}(2003){Lilly}, {Carollo}, \& {Stockton}}]{lilly03}
{Lilly}, S.~J., {Carollo}, C.~M., \& {Stockton}, A.~N. 2003, \apj, 597, 730

\bibitem[{{Lin} {et~al.}(2017){Lin}, {Lin}, {Hsu}, {et~al.}}]{lin17}
{Lin}, L., {Lin}, J.-H., {Hsu}, C.-H., {et~al.} 2017, arXiv:1702.02464

\bibitem[{{Liu} {et~al.}(2013){Liu}, {Zakamska}, {Greene}, {Nesvadba}, \&
  {Liu}}]{liu13}
{Liu}, G., {Zakamska}, N.~L., {Greene}, J.~E., {Nesvadba}, N.~P.~H., \& {Liu},
  X. 2013, \mnras, 430, 2327

\bibitem[{{Mannucci} {et~al.}(2010){Mannucci}, {Cresci}, {Maiolino}, {Marconi},
  \& {Gnerucci}}]{mannucci10}
{Mannucci}, F., {Cresci}, G., {Maiolino}, R., {Marconi}, A., \& {Gnerucci}, A.
  2010, \mnras, 408, 2115

\bibitem[{{Martin}(2005)}]{martin05}
{Martin}, C.~L. 2005, \apj, 621, 227

\bibitem[{{Martin} {et~al.}(2012){Martin}, {Shapley}, {Coil},
  {et~al.}}]{martin12}
{Martin}, C.~L., {Shapley}, A.~E., {Coil}, A.~L., {et~al.} 2012, \apj, 760, 127

\bibitem[{{McLean} {et~al.}(2012){McLean}, {Steidel}, {Epps},
  {et~al.}}]{mclean12}
{McLean}, I.~S., {Steidel}, C.~C., {Epps}, H.~W., {et~al.} 2012, in \procspie,
  Vol. 8446, Ground-based and Airborne Instrumentation for Astronomy IV, 84460J

\bibitem[{{Mutch} {et~al.}(2013){Mutch}, {Croton}, \& {Poole}}]{mutch13}
{Mutch}, S.~J., {Croton}, D.~J., \& {Poole}, G.~B. 2013, \mnras, 435, 2445

\bibitem[{{Nakajima} \& {Ouchi}(2014)}]{nakajima14}
{Nakajima}, K. \& {Ouchi}, M. 2014, \mnras, 442, 900

\bibitem[{{Nakajima} {et~al.}(2013){Nakajima}, {Ouchi}, {Shimasaku},
  {et~al.}}]{nakajima13}
{Nakajima}, K., {Ouchi}, M., {Shimasaku}, K., {et~al.} 2013, \apj, 769, 3

\bibitem[{{Nesvadba} {et~al.}(2008){Nesvadba}, {Lehnert}, {De Breuck},
  {Gilbert}, \& {van Breugel}}]{nesvadba08}
{Nesvadba}, N.~P.~H., {Lehnert}, M.~D., {De Breuck}, C., {Gilbert}, A.~M., \&
  {van Breugel}, W. 2008, \aap, 491, 407

\bibitem[{{Newman} {et~al.}(2012{\natexlab{a}}){Newman}, {Genzel},
  {F{\"o}rster-Schreiber}, {et~al.}}]{newman12b}
{Newman}, S.~F., {Genzel}, R., {F{\"o}rster-Schreiber}, N.~M., {et~al.}
  2012{\natexlab{a}}, \apj, 761, 43

\bibitem[{{Newman} {et~al.}(2012{\natexlab{b}}){Newman}, {Shapiro Griffin},
  {Genzel}, {et~al.}}]{newman12a}
{Newman}, S.~F., {Shapiro Griffin}, K., {Genzel}, R., {et~al.}
  2012{\natexlab{b}}, \apj, 752, 111

\bibitem[{{Noeske} {et~al.}(2007){Noeske}, {Weiner}, {Faber},
  {et~al.}}]{noeske07}
{Noeske}, K.~G., {Weiner}, B.~J., {Faber}, S.~M., {et~al.} 2007, \apjl, 660,
  L43

\bibitem[{{Oey} \& {Kennicutt}(1997)}]{oey97}
{Oey}, M.~S. \& {Kennicutt}, Jr., R.~C. 1997, \mnras, 291, 827

\bibitem[{{Oke} \& {Gunn}(1983)}]{oke83}
{Oke}, J.~B. \& {Gunn}, J.~E. 1983, \apj, 266, 713

\bibitem[{{Oppenheimer} {et~al.}(2010){Oppenheimer}, {Dav{\'e}}, {Kere{\v s}},
  {et~al.}}]{oppenheimer10}
{Oppenheimer}, B.~D., {Dav{\'e}}, R., {Kere{\v s}}, D., {et~al.} 2010, \mnras,
  406, 2325

\bibitem[{{Osterbrock}(1989)}]{osterbrock89}
{Osterbrock}, D.~E. 1989, \skytel, 78, 491

\bibitem[{{Pagel} {et~al.}(1979){Pagel}, {Edmunds}, {Blackwell}, {Chun}, \&
  {Smith}}]{pagel79}
{Pagel}, B.~E.~J., {Edmunds}, M.~G., {Blackwell}, D.~E., {Chun}, M.~S., \&
  {Smith}, G. 1979, \mnras, 189, 95

\bibitem[{{Pellegrini} {et~al.}(2012){Pellegrini}, {Oey}, {Winkler},
  {et~al.}}]{pellegrini12}
{Pellegrini}, E.~W., {Oey}, M.~S., {Winkler}, P.~F., {et~al.} 2012, \apj, 755,
  40

\bibitem[{{Prochaska} {et~al.}(2011){Prochaska}, {Kasen}, \&
  {Rubin}}]{prochaska11}
{Prochaska}, J.~X., {Kasen}, D., \& {Rubin}, K. 2011, \apj, 734, 24

\bibitem[{{Ranalli} {et~al.}(2016){Ranalli}, {Koulouridis}, {Georgantopoulos},
  {et~al.}}]{ranalli16}
{Ranalli}, P., {Koulouridis}, E., {Georgantopoulos}, I., {et~al.} 2016, \aap,
  590, A80

\bibitem[{{Rubin} {et~al.}(2014){Rubin}, {Prochaska}, {Koo},
  {et~al.}}]{rubin14}
{Rubin}, K.~H.~R., {Prochaska}, J.~X., {Koo}, D.~C., {et~al.} 2014, \apj, 794,
  156

\bibitem[{{Rupke} {et~al.}(2005{\natexlab{a}}){Rupke}, {Veilleux}, \&
  {Sanders}}]{rupke05a}
{Rupke}, D.~S., {Veilleux}, S., \& {Sanders}, D.~B. 2005{\natexlab{a}}, \apjs,
  160, 87

\bibitem[{{Rupke} {et~al.}(2005{\natexlab{b}}){Rupke}, {Veilleux}, \&
  {Sanders}}]{rupke05b}
---. 2005{\natexlab{b}}, \apjs, 160, 115

\bibitem[{{Rupke} {et~al.}(2017){Rupke}, {G{\"u}ltekin}, \&
  {Veilleux}}]{rupke17}
{Rupke}, D.~S.~N., {G{\"u}ltekin}, K., \& {Veilleux}, S. 2017, \apj, 850, 40

\bibitem[{{Salpeter}(1955)}]{salpeter55}
{Salpeter}, E.~E. 1955, ApJ, 121, 161

\bibitem[{{Schawinski} {et~al.}(2015){Schawinski}, {Koss}, {Berney}, \&
  {Sartori}}]{schawinski15}
{Schawinski}, K., {Koss}, M., {Berney}, S., \& {Sartori}, L.~F. 2015, \mnras,
  451, 2517

\bibitem[{{Schirmer} {et~al.}(2013){Schirmer}, {Diaz}, {Holhjem}, {Levenson},
  \& {Winge}}]{schirmer13}
{Schirmer}, M., {Diaz}, R., {Holhjem}, K., {Levenson}, N.~A., \& {Winge}, C.
  2013, \apj, 763, 60

\bibitem[{{Schirmer} {et~al.}(2016){Schirmer}, {Malhotra}, {Levenson},
  {et~al.}}]{schirmer16}
{Schirmer}, M., {Malhotra}, S., {Levenson}, N.~A., {et~al.} 2016, \mnras, 463,
  1554

\bibitem[{{Shields}(1990)}]{shields90}
{Shields}, G.~A. 1990, \araa, 28, 525

\bibitem[{{Shirazi} \& {Brinchmann}(2012)}]{shirazi12}
{Shirazi}, M. \& {Brinchmann}, J. 2012, \mnras, 421, 1043

\bibitem[{{Shirazi} {et~al.}(2014){Shirazi}, {Brinchmann}, \&
  {Rahmati}}]{shirazi14}
{Shirazi}, M., {Brinchmann}, J., \& {Rahmati}, A. 2014, \apj, 787, 120

\bibitem[{{Silk} \& {Rees}(1998)}]{silk98}
{Silk}, J. \& {Rees}, M.~J. 1998, \aap, 331, L1

\bibitem[{{Simpson} {et~al.}(2006){Simpson}, {Mart{\'{\i}}nez-Sansigre},
  {Rawlings}, {et~al.}}]{simpson06}
{Simpson}, C., {Mart{\'{\i}}nez-Sansigre}, A., {Rawlings}, S., {et~al.} 2006,
  \mnras, 372, 741

\bibitem[{{Simpson} {et~al.}(2012){Simpson}, {Rawlings}, {Ivison},
  {et~al.}}]{simpson12}
{Simpson}, C., {Rawlings}, S., {Ivison}, R., {et~al.} 2012, \mnras, 421, 3060

\bibitem[{{Somerville} {et~al.}(2008){Somerville}, {Hopkins}, {Cox},
  {Robertson}, \& {Hernquist}}]{somerville08}
{Somerville}, R.~S., {Hopkins}, P.~F., {Cox}, T.~J., {Robertson}, B.~E., \&
  {Hernquist}, L. 2008, \mnras, 391, 481

\bibitem[{{Soto} {et~al.}(2012){Soto}, {Martin}, {Prescott}, \&
  {Armus}}]{soto12}
{Soto}, K.~T., {Martin}, C.~L., {Prescott}, M.~K.~M., \& {Armus}, L. 2012,
  \apj, 757, 86

\bibitem[{{Steidel} {et~al.}(2010){Steidel}, {Erb}, {Shapley},
  {et~al.}}]{steidel10}
{Steidel}, C.~C., {Erb}, D.~K., {Shapley}, A.~E., {et~al.} 2010, \apj, 717, 289

\bibitem[{{Str{\"u}der} {et~al.}(2001){Str{\"u}der}, {Briel}, {Dennerl},
  {et~al.}}]{struder01}
{Str{\"u}der}, L., {Briel}, U., {Dennerl}, K., {et~al.} 2001, \aap, 365, L18

\bibitem[{{Sun} {et~al.}(2017){Sun}, {Greene}, \& {Zakamska}}]{sun17}
{Sun}, A.-L., {Greene}, J.~E., \& {Zakamska}, N.~L. 2017, \apj, 835, 222

\bibitem[{{Thuan} \& {Izotov}(2005)}]{thuan05}
{Thuan}, T.~X. \& {Izotov}, Y.~I. 2005, \apjs, 161, 240

\bibitem[{{Tremonti} {et~al.}(2004){Tremonti}, {Heckman}, {Kauffmann},
  {et~al.}}]{tremonti04}
{Tremonti}, C.~A., {Heckman}, T.~M., {Kauffmann}, G., {et~al.} 2004, \apj, 613,
  898

\bibitem[{{Turner} {et~al.}(2001){Turner}, {Abbey}, {Arnaud},
  {et~al.}}]{turner01}
{Turner}, M.~J.~L., {Abbey}, A., {Arnaud}, M., {et~al.} 2001, \aap, 365, L27

\bibitem[{{Ueda} {et~al.}(2008){Ueda}, {Watson}, {Stewart}, {et~al.}}]{ueda08}
{Ueda}, Y., {Watson}, M.~G., {Stewart}, I.~M., {et~al.} 2008, \apjs, 179, 124

\bibitem[{{Ulrich} {et~al.}(1997){Ulrich}, {Maraschi}, \& {Urry}}]{ulrich97}
{Ulrich}, M.-H., {Maraschi}, L., \& {Urry}, C.~M. 1997, \araa, 35, 445

\bibitem[{{van de Voort} {et~al.}(2011){van de Voort}, {Schaye}, {Booth}, \&
  {Dalla Vecchia}}]{vandevoort11}
{van de Voort}, F., {Schaye}, J., {Booth}, C.~M., \& {Dalla Vecchia}, C. 2011,
  \mnras, 415, 2782

\bibitem[{{Veilleux} \& {Osterbrock}(1987)}]{veilleux87}
{Veilleux}, S. \& {Osterbrock}, D.~E. 1987, \apjs, 63, 295

\bibitem[{{Weber} {et~al.}(2019){Weber}, {Pauldrach}, \& {Hoffmann}}]{weber19}
{Weber}, J.~A., {Pauldrach}, A.~W.~A., \& {Hoffmann}, T.~L. 2019, \aap, 622,
  A115

\bibitem[{{Weiner} {et~al.}(2009){Weiner}, {Coil}, {Prochaska},
  {et~al.}}]{weiner09}
{Weiner}, B.~J., {Coil}, A.~L., {Prochaska}, J.~X., {et~al.} 2009, \apj, 692,
  187

\bibitem[{{Yabe} {et~al.}(2009){Yabe}, {Ohta}, {Iwata}, {et~al.}}]{yabe09}
{Yabe}, K., {Ohta}, K., {Iwata}, I., {et~al.} 2009, \apj, 693, 507

\bibitem[{{Yang} {et~al.}(2017){Yang}, {Malhotra}, {Rhoads}, \&
  {Wang}}]{yang17}
{Yang}, H., {Malhotra}, S., {Rhoads}, J.~E., \& {Wang}, J. 2017, \apj, 847, 38

\bibitem[{{Yuma} {et~al.}(2013){Yuma}, {Ouchi}, {Drake}, {et~al.}}]{yuma13}
{Yuma}, S., {Ouchi}, M., {Drake}, A.~B., {et~al.} 2013, \apj, 779, 53

\bibitem[{{Yuma} {et~al.}(2017){Yuma}, {Ouchi}, {Drake}, {et~al.}}]{yuma17}
---. 2017, \apj, 841, 93

\bibitem[{{Zhang} {et~al.}(2017){Zhang}, {Yan}, {Bundy}, {et~al.}}]{zhang17}
{Zhang}, K., {Yan}, R., {Bundy}, K., {et~al.} 2017, \mnras, 466, 3217

\end{thebibliography}
\end{document}